%% file: paper.tex
\documentclass[sigplan,screen,nonacm]{acmart}

\copyrightyear{2026}
\acmYear{2026}
\setcopyright{cc}
\setcctype{by}
\acmConference[ASPLOS '26]{Proceedings of the 31st ACM International Conference on Architectural Support for Programming Languages and Operating Systems, Volume 2}{March 22--26, 2026}{Pittsburgh, PA, USA}
\acmBooktitle{Proceedings of the 31st ACM International Conference on Architectural Support for Programming Languages and Operating Systems, Volume 2 (ASPLOS '26), March 22--26, 2026, Pittsburgh, PA, USA}
\acmDOI{10.1145/3779212.3790138}
\acmISBN{979-8-4007-2359-9/2026/03}

\input{util/packages}

\input{util/macros}



\author{Bilel Sefsaf}
\orcid{0009-0007-9367-4678}
\affiliation{%
  \institution{New York University Abu Dhabi}
  \city{Abu Dhabi}
  \country{United Arab Emirates}
}
\affiliation{%
  \institution{Ecole Superieure d'Informatique}
  \city{Algiers}
  \country{Algeria}
}
\email{kb_sefsaf@esi.dz}

\author{Abderraouf Dandani}
\orcid{0009-0002-5301-4388}
\affiliation{%
  \institution{New York University Abu Dhabi}
  \city{Abu Dhabi}
  \country{United Arab Emirates}
}
\affiliation{%
  \institution{Ecole Superieure d'Informatique}
  \city{Algiers}
  \country{Algeria}
}
\email{ka_dandani@esi.dz}

\author{Abdessamed Seddiki}
\orcid{0009-0004-4452-0777}
\affiliation{%
  \institution{New York University Abu Dhabi}
  \city{Abu Dhabi}
  \country{United Arab Emirates}
}
\affiliation{%
  \institution{Ecole Superieure d'Informatique}
  \city{Algiers}
  \country{Algeria}
}
\email{ka_seddiki@esi.dz}

\author{Arab Mohammed}
\orcid{0009-0000-1471-0431}
\affiliation{%
  \institution{New York University Abu Dhabi}
  \city{Abu Dhabi}
  \country{United Arab Emirates}
}
\affiliation{%
  \institution{Ecole Superieure d'Informatique}
  \city{Algiers}
  \country{Algeria}
}
\email{km_arab@esi.dz}

\author{Eduardo Chielle}
\orcid{0000-0002-1938-912X}
\affiliation{%
  \department{Center for Cyber Security}
  \institution{New York University Abu Dhabi}
  \city{Abu Dhabi}
  \country{United Arab Emirates}
}
\email{ec126@nyu.edu}

\author{Michail Maniatakos}
\orcid{0000-0001-6899-0651}
\affiliation{%
  \department{Center for Cyber Security}
  \institution{New York University Abu Dhabi}
  \city{Abu Dhabi}
  \country{United Arab Emirates}
}
\email{mm6446@nyu.edu}

\author{Riyadh Baghdadi}
\orcid{0000-0002-9350-3998}
\affiliation{%
  \institution{New York University Abu Dhabi}
  \city{Abu Dhabi}
  \country{United Arab Emirates}
}
\email{rb4792@nyu.edu}

\usepackage{placeins}
\renewcommand{\shortauthors}{Bilel Sefsaf et al.}
\begin{document}

\title{CHEHAB RL: Learning to Optimize Fully Homomorphic Encryption Computations}

\begin{abstract}
    \input{0-abstract}
\end{abstract}

\maketitle 

\input{1-introduction}

\input{2.5_Motivating_Example}

\input{2-background}
\input{3-overview}

\input{4-optmization}
\input{5-training}
\input{6-evaluation}

\input{7-related_work}
\input{8-conclusion}

\clearpage
\appendix
\input{9-appendix}

\balance
\bibliographystyle{ACM-Reference-Format}
\bibliography{references}

\end{document}

%% file: util/packages.tex
\usepackage{algorithm}
\usepackage{float}
\usepackage{makecell}
\setcellgapes{3pt}
\usepackage{amsfonts}
\usepackage{xcolor}
\usepackage{soul}
\usepackage{multirow}
\usepackage{makecell}
\usepackage[unicode]{hyperref}
\usepackage{graphicx} 
\usepackage{subcaption}

\usepackage{listings}
\usepackage{url}

\lstdefinestyle{promptstyle}{
    basicstyle=\ttfamily\scriptsize\color{editColor},
    breaklines=true,               
    breakatwhitespace=true,
    columns=fullflexible,
    frame=single,
    backgroundcolor=\color{gray!3},
    commentstyle=\color{gray}\upshape,
    keywordstyle=\color{blue}\mdseries,
    stringstyle=\color{purple},
    keepspaces=true,
    literate={–}{{-}}1 {—}{{-}}1 {…}{...}1 {§}{{\S}}1
}

\lstdefinestyle{shell}{
    basicstyle=\small\ttfamily,
    breaklines=true,
    postbreak=\mbox{\textcolor{red}{$\hookrightarrow$}\space},
    language=bash,
    xleftmargin=1.5em,
    showstringspaces=false
}

\newcommand{\edit}[1]{{\normalfont\textcolor{editColor}{#1}}}
\usepackage{balance}

%% file: util/macros.tex
\newcommand{\shrink}[1][1]{\vspace{-#1\dimexpr0.25cm\relax}}

\definecolor{listinggreen}{rgb}{0,0.6,0}
\definecolor{listinggray}{rgb}{0.5,0.5,0.5}
\definecolor{listingmauve}{rgb}{0.58,0,0.82}
\definecolor{listingkeywordcolor}{rgb}{1.0,0.4,0.0}
\definecolor{listinglightgray}{rgb}{0.8863,0.8863,0.8863}
\definecolor{codegreen}{rgb}{0,0.6,0}
\definecolor{codegray}{rgb}{0.5,0.5,0.5}
\definecolor{codeblue}{rgb}{0,0,0.8}
\colorlet{editColor}{black} 
\newcommand{\yes}{\color{listinggreen}{\textbf{\scriptsize \checkmark}}}
\newcommand{\no}{\color{listingkeywordcolor}{\textbf{\scriptsize $\times$}}}

\newcommand{\h}[1]{\textbf{#1}} 

\newcommand{\Tokenization}{Identifier and Constant Invariant} 
\newcommand{\Token}{ICI}  

\newcommand{\fontsizefortables}{\scriptsize}

\newcommand{\CodeSize}{\footnotesize}
\newcommand{\SmallCodeSize}{\scriptsize}

\newcommand{\ExecutionTimeSpeedupOverCoyote}{5.3}
\newcommand{\CompilationTimeSpeedupOverCoyote}{27.9}
\newcommand{\NoiseReductionOverCoyote}{2.54}

\newif\ifshowextra
\showextratrue

%% file: 0-abstract.tex
Fully Homomorphic Encryption (FHE) enables computations directly on encrypted data, but its high computational cost remains a significant barrier. Writing efficient FHE code is a complex task requiring cryptographic expertise, and finding the optimal sequence of program transformations is often intractable. In this paper, we propose CHEHAB RL, a novel framework that leverages deep reinforcement learning (RL) to automate FHE code optimization. Instead of relying on predefined heuristics or combinatorial search, our method trains an RL agent to learn an effective policy for applying a sequence of rewriting rules to automatically vectorize scalar FHE code while reducing instruction latency and noise growth. The proposed approach supports the optimization of both structured and unstructured code.
To train the agent, we synthesize a diverse dataset of computations using a large language model (LLM).
We integrate our proposed approach into the CHEHAB FHE compiler and evaluate it on a suite of benchmarks, comparing its performance against Coyote, a state-of-the-art vectorizing FHE compiler. The results show that our approach generates code that is $\ExecutionTimeSpeedupOverCoyote{}\times$ faster in execution, accumulates $\NoiseReductionOverCoyote{}\times$ less noise, while the compilation process itself is $\CompilationTimeSpeedupOverCoyote{}\times$ faster than Coyote (geometric means).

%% file: 1-introduction.tex
\shrink
\section{Introduction}

In Fully Homomorphic Encryption (FHE), computations are performed directly on encrypted data without decryption. This enables a client to transmit data to a third party that processes it and returns the result without performing decryption. As a result, sensitive information remains protected.

Despite notable recent progress and increasing practical adoption~\cite{Viand2022HECOFH,MSR2021PasswordMonitor,gursoy2022privacy,KIM20211108,ICO2023HECaseStudy}, FHE remains computationally expensive. For example, a ciphertext\footnote{Ciphertext: encrypted data; Plaintext: clear data.} multiplication takes $10^6$ times longer than its plaintext counterpart\footnote{Evaluation done on the BFV scheme, for 128-bit security, using Microsoft SEAL~\cite{sealcrypto} and running on a modern multicore CPU.}.
This disparity is due to the inherent complexity of homomorphic operations, as a single ciphertext multiplication entails multiple n-order polynomial multiplications. Reducing the execution time of FHE applications is essential for their practical adoption. Consequently, experts devote substantial effort to optimizing FHE implementations.

Writing efficient code for FHE is tedious, though. It requires a considerable amount of expertise in writing low-level code for the target scheme. This makes writing code for FHE applications time-consuming and error-prone and hinders its wide adoption~\cite{Viand2022HECOFH,Chowdhary2021EVA}.
To address this problem, the research community has explored the development of compilers for FHE applications \cite{Viand2022HECOFH, Malik2023Coyote, Cowan2021Porcupine, Dathathri2020, Chowdhary2021EVA, Lee2022, Dathathri2019CHET}. HECO~\cite{Viand2022HECOFH}, for example, focuses on efficiently transforming structured code into the FHE paradigm with SIMD instructions (structured code, a.k.a. regular code, stands for code with loops). In a similar way, CHET~\cite{Dathathri2019CHET} focuses on the vectorization of structured computations over packed tensors (such as neural networks).
However, neither of these compilers supports vectorizing unstructured code (arbitrary, non-loop-based code).

More recent work, such as Coyote~\cite{Malik2023Coyote} and Porcupine~\cite{Cowan2021Porcupine}, proposes compilers that support vectorization of both structured and unstructured code. However, both Coyote and Porcupine frame vectorization as search over a combinatorial design space: which subexpressions to pack, how to lay out data in wide ciphertext vectors, where to insert rotations and masks, etc. In such a complex space, local choices have non-local effects (e.g., a packing decision can increase later rotations or depth), so simple heuristics or greedy rule application often get trapped in poor local optima. To mitigate this, Coyote couples hand-tuned heuristics with an ILP solver to select packs, while Porcupine uses program synthesis to enumerate and check candidates~\cite{Malik2023Coyote, Cowan2021Porcupine}. These strategies are effective, but they are limited in their scalability: exploring or solving over large candidate sets becomes computationally expensive as programs grow in size.

In this paper, we argue that Reinforcement Learning (RL) provides a better solution: we treat FHE code optimization as sequential decision-making and use RL to learn a policy that composes sequences of rewriting rules to minimize latency and noise. By optimizing a single global cost, the RL policy makes global decisions without explicit combinatorial enumeration or ILP, yielding high-quality transformations with substantially lower compile time.


More concretely, we propose CHEHAB RL, a novel framework that leverages reinforcement learning (RL) to automatically optimize code for the field of FHE. It supports both structured and unstructured code and scales better than existing approaches. CHEHAB RL automatically vectorizes code while minimizing noise growth and circuit latency. This is achieved through a policy network that selects a sequence of rewriting rules to apply to code to vectorize it while minimizing rotations, circuit depth, multiplicative depth, and other FHE operations.
Because our approach selects the sequence of rewriting rules using a policy network, it avoids computationally expensive search methods used in state-of-the-art FHE compilers, and therefore scales better.

\edit{Bringing RL into this setting is nontrivial and requires addressing several FHE-specific challenges:} \edit{1) Large action space. The agent chooses from an extensive set of rules, each applicable to multiple locations. This makes the action space large. To address this challenge, we propose a multi-discrete hierarchical policy supported by a position network where the agent first decides about which transformation to apply, then decides about where to apply it. We show that this design outperforms a flat action space.}
\edit{ 2) Fast, FHE-aware reward. Running FHE code during training to obtain a reward is prohibitively slow. To address this challenge, we introduce an analytical reward function tailored to the FHE domain. It combines operation cost, circuit depth, and multiplicative depth and captures both performance and noise growth, while enabling fast training.}
\edit{3) LLM-Synthesized dataset. Since no dataset of optimizable FHE programs exists, we build an LLM-based synthesis pipeline and use it to generate data and show that training on this data yields much better agents than training on random programs.}

We integrate our proposed approach into the CHEHAB FHE compiler and evaluate it on a set of kernels, comparing it to Coyote, a state-of-the-art FHE compiler, and demonstrate that it produces higher-quality code and scales better.

The contributions of the paper are as follows:
\begin{enumerate}
    \item We formulate the problem of FHE code optimization as a sequential decision-making problem, and propose to learn a policy using RL to solve it.
    \item We propose CHEHAB RL, a novel framework that leverages RL to automatically vectorize structured and unstructured FHE code and minimize noise growth and circuit latency while scaling better compared to state-of-the-art. To the best of our knowledge, this is the first use of RL to solve this problem.
    \item \edit{We use an LLM-guided synthesis pipeline to generate a large training dataset of FHE expressions (15,855 expressions), and show that training on this dataset outperforms training on randomly generated FHE expressions.}
    \item We compare CHEHAB RL to Coyote and show that it generates code that is $\ExecutionTimeSpeedupOverCoyote{}\times$ faster, accumulates $\NoiseReductionOverCoyote{}\times$ less noise, while it takes $\CompilationTimeSpeedupOverCoyote{}\times$ less time in compilation.
    \item We release CHEHAB RL to the community \footnote{\url{https://github.com/Modern-Compilers-Lab/CHEHAB}}.
\end{enumerate}

%% file: 2.5_Motivating_Example.tex
\section{Motivating Example\label{sec:motivatingexample}}

Vectorization in FHE is challenging as it introduces rotations, masking, and data packing. Poor vectorization can increase the number of rotations and circuit depth, severely impacting performance and increasing noise in the circuit. Thus, effective FHE vectorization must maximize code vectorization while minimizing rotation overhead and circuit depth.

Consider this unstructured (non-loop-based) code:

{
\CodeSize
\begin{equation}
x =  ( ((v_1 \cdot v_2) \cdot (v_3 \cdot v_4)) + ((v_3 \cdot v_4) \cdot (v_5 \cdot v_6)) )\cdot ((v_7 \cdot v_8) \cdot (v_9 \cdot v_{10}))
\label{eq:example}
\tag{1}
\end{equation}
}

\begin{figure}[h!t]
    \centering
    \includegraphics[width=0.55\linewidth]{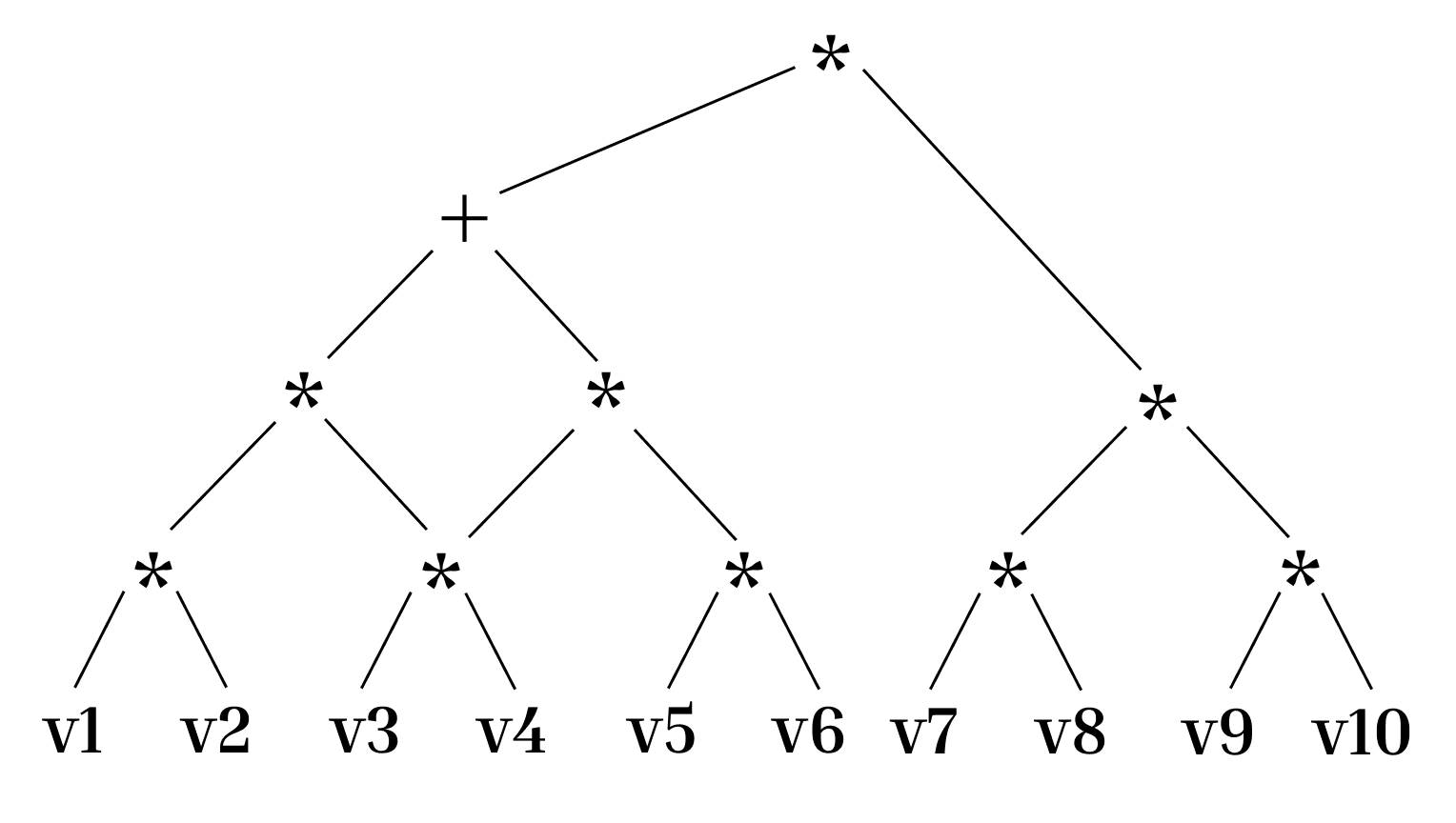}
    \shrink[1.5]
    \caption{Example of an unstructured scalar code.}
    \Description{A code snippet demonstrating unstructured scalar operations.}
    \shrink[2.4]
    \label{fig:motivatingExample}
\end{figure}

Fig.~\ref{fig:motivatingExample} represents the code as a circuit (a common representation of code in the FHE literature). This expression is not trivial to vectorize.

To vectorize it, we use a Term Rewriting System (TRS) consisting of a set of rules. For illustration, we show only three rules, but the TRS has more.

\begin{figure}[h!t]
    \centering
    \SmallCodeSize
    \begin{subfigure}{0.22\textwidth}
        \begin{minipage}{\linewidth}
            \centering
            \[
            \begin{aligned}
               & [1 , 2] = [v_1 , v_5] \cdot [v_2 , v_6] \\
               & [3 , \_] = [1 , \_] + [2 , \_] \\
               & [4 , \_] = [v_3 , \_] \cdot [v_4 , \_] \\
               & [5 , \_] = [3 , \_] \cdot [4 , \_] \\
               & [6 , 7] = [v_7 , v_9] \cdot [v_8 , v_{10}] \\
               & [8 , \_] = [6 , \_] \cdot [7 , \_] \\
               & [x , \_] = [5 , \_] \cdot [8 , \_]  \\          
            \end{aligned}
            \]
        \end{minipage}
        \caption{First approach.}
        \label{fig:vectorization1}
    \end{subfigure}
    \hfill
    \begin{subfigure}{0.22\textwidth}
        \begin{minipage}{\linewidth}
            \centering
            \[
            \begin{aligned}
              & [1 , 2 , 3] = [v_1 , v_5 , v_6] \cdot [v_2 , 1 , 1]  \\
              & [4 , \_ , \_] = [2, \_, \_] \cdot [3 , \_ , \_] \\
              & [5 , \_ , \_] = [1 , \_ , \_] + [4 , \_ , \_] \\
              & [6, \_, \_] = [v_3 , \_ , \_] \cdot [v_4 , \_ , \_] \\
              & [7 , \_ , \_] = [5 , \_ , \_] \cdot [6 , \_ , \_] \\
              & [8 , 9 , \_] = [v_7 , v_9 , \_] \cdot [v_8, v_{10} , \_] \\
              & [10 , \_ , \_] = [8 , \_ , \_] \cdot [9 , \_ , \_] \\
              & [x , \_ , \_] = [7 , \_ , \_] \cdot [10 , \_ , \_] \\
            \end{aligned}
            \]
        \end{minipage}
        \caption{Second approach.}
        \label{fig:vectorization2}
    \end{subfigure}
    \caption{Vectorization approaches.}
    \label{fig:vectorization}
    \shrink[1.5]
    \Description{Two listings comparing vectorization strategies. The left listing shows operations packed into 2-element vectors. The right listing shows a similar sequence packed into 3-element vectors with some padding.}
\end{figure}

{
\CodeSize
\(
\texttt{R1: multiplication\_commutativity},\quad a \cdot b \Rightarrow b \cdot a 
\)

\(
\texttt{R2: comm\_factor},\quad a \cdot b + a \cdot c \Rightarrow a \cdot (b + c)
\)

\(
\texttt{R3: vectorization},\quad (a \cdot b) + (c \cdot d) \Rightarrow \texttt{Vec}(a , c) \cdot \texttt{Vec}(b , d)
\)
}

We first apply the rules \texttt{R1} then \texttt{R2}, yielding:

{
\CodeSize
\begin{equation}
x = (  ( v_3 \cdot v_4) \cdot ((v_1 \cdot v_2) + (v_5 \cdot v_6))) \cdot ((v_7 \cdot v_8) \cdot (v_9 \cdot v_{10}))
\label{eq:vectorization-alt}
\tag{2}
\end{equation}
}

From this transformed expression, we can apply other rewriting rules to vectorize the code. There are many ways to vectorize it (also called vectorization strategies). We show two examples in Fig. \ref{fig:vectorization} (intermediate values are named using numbers). Each vectorization strategy is obtained by applying a different sequence of rewriting rules. 
The key challenge is determining the sequence of rewriting rules that yields the best vectorization strategy while minimizing the number of rotations and depth of the circuit (multiplicative depth, more precisely, as we illustrate later).

We omit showing the rotation and masking instructions in the example for brevity. For example, in Fig. \ref{fig:vectorization1}, in order to calculate the operands of the 2nd statement ($[1,\_]$ and $[2,\_]$), rotations and masking should be applied on the result of the 1st statement as follows\footnote{We assume 4-bit numbers in the example for simplicity, and we use \_ to denote a zero-valued slot.}:

{
\CodeSize
\[
\begin{aligned}
& [1,\_] = [1,2] \,\,\&\,\, [F,\_] \\
& [2,\_] = ([1,2]<<1) \,\,\&\,\, [F,\_]
\end{aligned}
\]
}

\noindent but we omit showing these rotations and maskings.


Let us compare the two vectorizations in Fig.~\ref{fig:vectorization} using an approximate cost model that assigns a relative latency to each operation: multiplications and rotations have a latency of 1, while additions have a latency of 0.1 (these values are toy values chosen for simplicity only, to illustrate the example). The original scalar expression has 9 multiplications and 1 addition, yielding a total cost of 9.1.
The vectorization in Fig.~\ref{fig:vectorization1} reduces the number of multiplications to 6 and additions to 1 but introduces 2 rotations, resulting in a total cost of 8.1. In contrast, the vectorization in Fig.~\ref{fig:vectorization2} involves 7 multiplications, 3 rotations, and 1 addition, yielding a 10.1 cost.

These results highlight that not all vectorizations are equally beneficial.
Moreover, applying \texttt{R1} prior to \texttt{R2} is essential to obtain the expression in Eq.~\ref{eq:vectorization-alt}, which can then be vectorized by applying \texttt{R3}. In contrast, starting with \texttt{R2} followed by \texttt{R1} yields Eq.~\ref{eq:example}, where \texttt{R3} is no longer applicable, thereby eliminating vectorization opportunities.
This limitation highlights the importance of carefully selecting which rules to apply and in which order.
To address this, we employ a policy network, trained via RL. The policy determines an ordered sequence of rules that optimize the code to achieve the best global vectorization, rather than relying on local improvements alone.

\shrink[1.5]
\paragraph{Vectorization in FHE}
Vectorization in FHE is different from classical vectorization. While many code vectorization methods have been studied in the literature~\cite{5260526,10.5555/502981,10.1145/1379022.1375595,10.5555/3314872.3314896,slp,goSLP,veGen}, such methods are not well-suited for vectorizing FHE code. First, since FHE does not support loops or conditionals, computations must be expressed as arithmetic circuits (unstructured code), preventing the direct use of classical loop-based vectorization methods.
FHE vectorization differs from classical SLP-based approaches (Superword-level parallelism)~\cite{slp,goSLP,veGen} due to the large vector sizes in FHE (thousands of slots vs a few in hardware registers) and the inability to index ciphertext vectors directly. Instead, expensive rotation operations simulate indexed accesses. In addition, rotations in FHE are computationally expensive and increase noise in the circuit, so vectorization strategies must minimize their use. These differences make classical SLP-based methods unsuitable for FHE, as we discuss in Sec.~\ref{sec:realted}.
Because ciphertext vectors are wider in FHE compared to traditional hardware vector registers, FHE vectorization yields significant speedups because a single ciphertext operation (add or multiply) applies element-wise to every packed slot in the ciphertext, so one operation can replace thousands of scalar steps, far beyond the 4–8-lane limits of hardware SIMD.

%% file: 2-background.tex
\shrink
\section{Background}

\subsection{Fully Homomorphic Encryption}

FHE allows computations on encrypted data directly (i.e., without decrypting the data). For a function \(f\) and a message \(m\), FHE allows the possibility of computing \(\widehat{y} = f(\widehat{m})\), without knowing \(m\).
Here, $\widehat{m}$ represents the encryption of $m$, and $\widehat{y}$ is the encrypted output, which must be decrypted by the owner of $m$ to obtain the output $\widehat{y} \mapsto y$ in plaintext.
FHE hence helps to achieve privacy-preserving computation in situations where sensitive data needs to be shared with third parties for computation offloading.


The Brakerski/Fan-Vercauteren (BFV) encryption scheme \cite{bfv_b,bfv_a:2012/144} is an FHE scheme that provides primitives for modular addition and multiplication along with primitives for noise management and ciphertext maintenance (i.e., relinearization, modulus switching, and bootstrapping).
BFV operates within plaintext and ciphertext spaces.
Plaintexts and ciphertexts are large order polynomials defined by the parameters $\{ n, t, q \}$.
The degree $n$ corresponds to the polynomial modulus $x^n + 1$, while the parameters $t$ and $q$ are the plaintext modulus and ciphertext modulus.


Encryption introduces \textit{noise} for security, which grows with operations on ciphertexts or between a ciphertext and a plaintext.
Its growth is limited by \( q \) and \( t \)\ifshowextra \ (defined in \ref{bfvapp})\fi; exceeding this limit corrupts the data, leading to incorrect decryption (a user sets a noise budget for their circuit, and the circuit has to run without exceeding that noise budget). \textit{Bootstrapping} can reset noise but has high computational costs.
Instead, FHE applications typically limit operations to a predefined depth.
Supporting deeper circuits requires increasing \( q \) and adjusting \( n \) for security.
This is especially important for circuits with high multiplicative depth, as multiplications cause exponential noise growth, while additions increase it linearly. 
However, larger \( q \) and \( n \) also introduce performance overheads due to larger polynomials and coefficients. Therefore, one of the goals when optimizing an FHE circuit is to minimize the multiplicative depth to be able to use the smallest \( q \) and \( n \) that still respect the noise budget in the circuit. This, in turn, makes FHE operations run faster and reduces the overall circuit latency.


Vectorization in FHE (a.k.a., \textit{batching}) is a technique that encodes a vector of $n$ integers in a single plaintext polynomial using the Chinese Remainder Theorem (CRT). Vectors in FHE are large (thousands of slots).
With vectorization, additions and multiplications are executed slot-wise in a Single-Instruction Multiple-Data (SIMD) fashion.
Moreover, BFV supports rotations of the encoded vector to move data (and perform data packing).
Rotation as a function takes a ciphertext (vector) and a rotation step and returns the ciphertext (vector) with slots shifted in a cyclic manner. For example, the rotation $[1,2,3]<<1$ returns $[2,3,1]$.
Note that FHE supports only additions, multiplications, and rotations and does not support branches or loops.
\ifshowextra 
See Appendix~\ref{appendix:fhe} for a formal definition of FHE concepts.
\fi 

\subsubsection{Circuit Depth and Multiplicative Depth\label{sec:depth}}

\shrink[0.75]
\paragraph{Circuit Depth.}
An FHE program can also be represented as a circuit. A circuit is a DAG (Directed Acyclic Graph) representation of the program that captures its dataflow. Calculating the depth of the circuit is important for estimating noise.
The depth of a circuit (DAG) represents the maximum number of consecutive operations performed on an input to compute the output. The depth of a node $v$ in the DAG is defined as the length of the longest path between $v$ and an input node.
This depth is usually used to quantify the noise in the circuit. However, it does not provide any information on the type of operations involved (which is necessary to determine how the noise grows). Since multiplication has the most significant impact on noise, one should compute the multiplicative depth of the circuit to estimate noise better.

\shrink[0.5]
\paragraph{Multiplicative Depth of a Circuit.}
Similar to circuit depth, except that we consider counting the number of multiplications only, since multiplications increase noise exponentially.
The multiplicative depth of a node $v$ is defined as the length of the longest path, when counting multiplication operations alone, between $v$ and an input node.






%% file: 3-overview.tex
\shrink[1]
\section{CHEHAB RL Overview}

\begin{figure}[t]
    \shrink
    \centering
    \includegraphics[width=\linewidth]{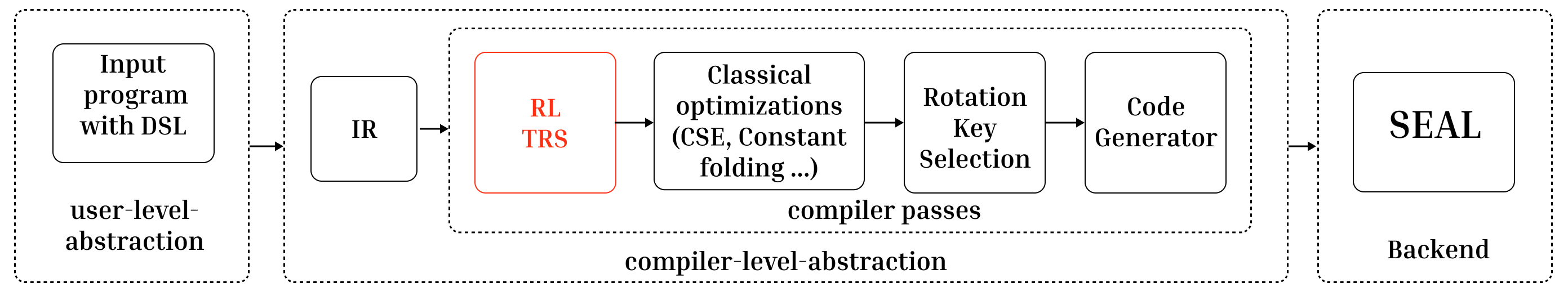}
    \Description{A high-level diagram of the CHEHAB compiler framework.}
    \caption{CHEHAB RL overview (our contribution: \emph{RL TRS}).}
    \label{fig:overview}
\end{figure}

This section provides an overview of CHEHAB RL and its components. We first show the overall design of the CHEHAB compiler and then discuss its major components, including our RL agent for code optimization.
CHEHAB defines a domain-specific language (DSL) for FHE. The CHEHAB DSL is the compiler's starting point. An intermediate representation (IR) is first generated from the CHEHAB DSL. This IR is then optimized by a Reinforcement learning (RL) agent, which learns a policy to apply a sequence of transformations. These transformations are drawn from a set of rewriting rules that include rules for both code vectorization and for simplification to reduce noise growth and instruction latency (i.e., rewrite complex arithmetic expressions into simpler expressions that have a smaller number of operations and depth).
Next, the compiler performs the rotation key selection. This pass generates the necessary rotation keys for rotation operations found in the optimized code.
Finally, the code generator generates a sequence of vectorized operations implemented in C++ and targeting Microsoft SEAL’s backend ~\cite{sealcrypto} for BFV. 
\edit{
The key new contribution introduced by this paper is CHEHAB RL, the RL-guided term rewriting system, which we integrate into the CHEHAB compiler as shown in Fig.~\ref{fig:overview}}.
\ifshowextra 
We present the RL agent in detail in the paper, while we present the rotation key selection method in Appendix \ref{appendix:RotationKeysSelection}.
\fi



\shrink[1]
\subsection{CHEHAB Domain Specific Language (DSL)}\label{subsec:DSL}

FHE performance and correctness critically depend on managing the encryption parameters and rotation keys.
Managing these keys and parameters manually is a time-consuming task that requires expertise. The CHEHAB DSL is designed to manage this complexity automatically, and thus improves the productivity of the developer. The developer expresses their code using a simple, high-level C++ code, while CHEHAB takes care of lowering that code to the FHE paradigm and takes care of generating and managing all the FHE-related objects (such as the encryption keys).
CHEHAB is embedded in C++, allowing users to leverage C++ features (e.g., templates) while extending it by defining the Plaintext and Ciphertext types. It also overloads basic operations for these types, mapping them to native FHE operations. One of the guiding principles in designing CHEHAB is the clear separation between computations and implementation details. In particular, it allows computations to be specified independently from encryption parameters, offering greater flexibility in exploiting potential optimization opportunities.

A program written in CHEHAB has three parts: (1) Input declaration: This section defines the inputs of the program. An input can be either a ciphertext or a plaintext. It can also be either a scalar or a vector. (2) Computations: This section specifies the computations, whether arithmetic or rotation operations, using standard C++ operators (\texttt{+}, \texttt{-}, \texttt{*}, \texttt{<<}). These operators are mapped to either vector or scalar operators depending on the type of their operands. (3) Outputs: This section declares the outputs.

The following example shows how to express the motivating example in our DSL. The list of operations supported by CHEHAB is presented in Table \ref{tab:operations} in the appendix.

{
    \CodeSize
    \begin{lstlisting}[
        language=C++,
        xleftmargin=-14pt,
        frame=none,
        framesep=0pt,
        columns=fullflexible,
        basicstyle=\ttfamily,
        keywordstyle=\color{codeblue}\bfseries,
        commentstyle=\color{codegreen},  
        stringstyle=\color{purple},  
        escapeinside={||},
        mathescape=true
    ]
    Ciphertext motivating_example() {
      Ciphertext v1,v2,v3,v4,v5,v6,v7,v8,v9,v10; //Inputs
      Ciphertext x =  (((v1 * v2) * (v3 * v4)) + ((v3 * v4) *
      (v5 * v6))) * ((v7 * v8) * (v9 * v10)); //Computations
      x.set_output(); //Outputs
      return x;
    }
    \end{lstlisting}
    \shrink[1]
}

\shrink
\subsection{CHEHAB Intermediate Representation (IR)}
The DSL program is compiled into an intermediate representation (IR) in the form of an AST (Abstract Syntax Tree), as is commonly done in compilers. Fig.~\ref{fig:motivatingExample} shows an example of an AST representation of the motivating example. From this representation, we extract a dataflow graph representation (DAG) of the code when needed (e.g., to calculate the multiplicative depth). To lower the DSL to the IR, we follow the same approach of Halide~\cite{10.1145/3150211} and Tiramisu~\cite{10.5555/3314872.3314896}. The approach relies on the use of templates and staging in C++.

\shrink[1]
\subsection{Code Optimizations}

After generating the IR, the next stage is code optimization. CHEHAB RL uses an RL agent to optimize code. The agent chooses a sequence of rules from its rule set. This set includes rules for \emph{vectorization}, which group scalar operations into vector ones, and rules for \emph{simplification}, which minimize noise growth and latency in the circuit. To further optimize code, the original CHEHAB compiler applies classic compiler optimizations, including common subexpression elimination (CSE), constant folding, and dead code elimination.

\shrink[1]
\subsection{Code generation}

In the final stage, CHEHAB generates code that targets the SEAL API~\cite{sealcrypto} by mapping each node in the IR (operator) to its equivalent API call in SEAL. During code generation, we also generate the necessary masking operations, private keys, public keys, and relinearization keys. \ifshowextra More details about code generation in Appendix~\ref{appendix:codegen}.\fi

%% file: 4-optmization.tex
\section{FHE Circuit Optimization} \label{sec:circuit_opt}
The core of the CHEHAB RL framework is a Term Rewriting System (TRS) that uses RL to transform an initial Intermediate Representation (IR) into an efficient one. We frame the optimization task as a sequential decision-making problem, formally a Markov Decision Process (MDP).


Our RL framework uses the actor-critic RL architecture and is defined by the following key components: 1) \emph{State representation (detailed in Sec. \ref{state}):} provides a representation of the program being optimized;
2) \emph{Action space (Sec. \ref{action}):} defines the set of rewriting rules that the agent can apply to the program being optimized;
3) \emph{Reward model (Sec. \ref{reward}):} provides a function that scores the quality of actions, guiding the agent toward actions that produce FHE circuits that are more efficient and have less noise;
4) \emph{Actor-critic networks (Sec. \ref{sec:actor-critic}):} the actor selects the next rewriting rule to apply given the current program state, and selects where to apply it, while the critic estimates the expected return and provides the learning signal for policy optimization during training (the critic is only used during training to train the policy).

We train our proposed RL agent on an \emph{LLM-generated dataset} (described in Sec. \ref{data}) using the \emph{PPO learning algorithm} (Sec. \ref{sec:agent_training}). The following sections describe these components.

\subsection{State Representation\label{state}}

The goal of the state representation is to produce a fixed-length vector embedding, $s_P$, that represents a given program IR, $P$. 
To extract the program representation, we first tokenize the program being optimized. We use a simple method that we call \emph{\Tokenization{} (\Token{})} tokenization.
\edit{
Conceptually, ICI is a form of \emph{alpha-renaming}~\cite{aho2006compilers} combined with simple program canonicalization.
ICI produces a canonical token sequence that is invariant to identifier names and to the concrete values of constants. For each program P, we rename variables such that the first distinct variable is mapped to v0, the second to v1, and so on. 
For example, the two semantically equivalent expressions \texttt{(+ a (+ b c))} and \texttt{(+ x (+ y z))} are both mapped to the exact same canonical sequence: \texttt{(+ v0 (+ v1 v2))}.
This normalization has a few practical benefits for learning:
(i) it collapses programs that differ only by identifier naming or by value-agnostic constants into the same representation, reducing the effective vocabulary and simplifying generalization; (ii) it produces a stable canonical form that we reuse for dataset deduplication, avoiding redundant near-duplicate training samples; (iii) ICI tokenization is computationally lightweight: it is a single linear scan with $O(1)$ expected-time hash map operations per new symbol. This contrasts with learned subword tokenizers such as Byte Pair Encoding (BPE)~\cite{sennrich-etal-2016-neural}, which repeatedly applies merge rules and performs large-vocabulary prefix searches, adding nontrivial overhead.
}

\edit{
More precisely, ICI uses a small fixed vocabulary for IR operators (e.g., \texttt{+}, \texttt{*}, \texttt{VecAdd}) and delimiters/parentheses, and builds per-program hash maps for identifiers and constants in a single
left-to-right pass. For each program $P$, the first distinct variable is mapped to \texttt{v0},
the second to \texttt{v1}, and so on. Numerical constants are mapped similarly to \texttt{c0},
\texttt{c1}, \ldots with one important exception: the integers \textbf{0} and \textbf{1} are kept as literal tokens because they play a special semantic role in many rewrite rules (additive/multiplicative identities). For all other constants, we discard the literal value but preserve (i) that the token is a constant and (ii) whether two constant occurrences in the same expression are equal (e.g., the same constant reused in multiple positions receives the same \texttt{c\#} token). This is sufficient for our term-rewriting system because our rewrite rules do not branch on specific numeric values beyond special cases such as 0/1. In practice, programs that differ only in identifier names or non-(0/1) constant values are typically optimized by the same rewrite strategy.
}

After tokenization, the program $P$ is represented by a variable-length sequence of tokens, $T(P) = [t_1, t_2, \dots, t_k]$. We then convert this sequence into a fixed-length vector that captures the long-range dependencies and hierarchical structure implied by the tokens. For this, we use a Transformer encoder~\cite{vaswani2017attention} that learns automatically a short, fixed-length embedding for the tokenized program. We chose this architecture because its self-attention mechanism is designed to model long-distance dependencies inherent in source code, a capability validated by state-of-the-art code models such as CodeBERT~\cite{feng2020codebert}. This is critical for processing the long token sequences that represent unrolled FHE programs. Early in the project, we compared the Transformer architecture to recurrent architectures, including LSTMs and GRUs, but they did not provide satisfactory results.
The Transformer encoder follows a standard architecture, consisting of a stack of 4 identical layers. Each layer employs a multi-head attention mechanism with 8 attention heads. The input token embeddings are summed with absolute positional encodings to preserve sequence order, and the encoder processes this sequence to generate a 256-dimensional vector for each token (we use the CLS special token as a placeholder to summarize the entire input sequence, as is commonly done in the literature~\cite{devlin2019bert}).




\shrink[1.5]
\subsection{Action Space \label{action}}

The action space defines the set of valid transformations the agent can perform on the current program state. In our framework, an action corresponds to a rewriting rule selected from a set of predefined rewriting rules.

A challenge in designing the action space is handling the location of rule application. A single rule might match multiple sub-expressions within the IR. To address this, the agent needs to select both a rule and a location. To enable this, we propose a hierarchical action space composed of two steps: the agent first picks a rule to apply, then picks the location where to apply it. To implement this, we propose a \edit{\emph{rule selection network}} used to pick an action, and a \emph{location \edit{selection} network} used to pick a location. \ifshowextra\edit{We compare this hierarchical formulation to a flat rule/location action space in Sec.~\ref{ablation:flat_vs_heir}, and show that the hierarchical agent learns faster and achieves higher average rewards in training compared to a flat action space (Fig.~\ref{fig:flat_vs_hier}).}\fi




To allow the agent to control the length of the optimization process, the action space also includes a special \emph{END} action. Selecting this action terminates the current optimization episode, allowing the agent to learn not only which rules to apply, but also when to stop.

In total, our TRS consists of 84 rewrite rules, in addition to the \texttt{END} action. To design these rules, we began by collecting rules from Halide's TRS~\cite{10.1145/3150211,Newcomb2020}, and excluded rules that are not compatible with FHE (comparison, division, etc.). We then added new rules designed to reduce circuit depth, multiplicative depth, rotations, and the total number of operations.
\ifshowextra
More details about the rewriting rules in Appendix \ref{rules}.
\fi

\subsection{Reward Model \label{reward}}
The agent learns via a reward signal designed to guide it toward producing FHE circuits that have minimal latency and noise. This signal is derived from an FHE-aware cost function that evaluates the quality of a given IR. We will first show the cost function, and then show how it is used to compute the reward.

\subsubsection{FHE Cost Function}\label{reward_cost}
The goal of this cost function is to capture the metrics that we aim to minimize: number of FHE operations, multiplicative depth, and depth of the circuit. We also want to minimize the use of scalar operations in favor of vector operations. By minimizing the previous metrics, we minimize latency and noise. To achieve our goal, we define the cost function of an IR expression, \emph{e}, as the weighted sum of the metrics that we want to minimize:

{
\CodeSize
\shrink[1.5]
\begin{align*}
\text{Cost}(\text{e}) = w_{\text{ops}} \cdot C_{\text{ops}}(\text{e}) 
                      + w_{\text{depth}} \cdot D_{\text{circuit}}(\text{e})
                      + w_{\text{mult}} \cdot D_{\text{mult}}(\text{e})   
\end{align*}
}

Where $C_{\text{ops}}$ is the cost of operations, $D_{\text{circuit}}$ is its circuit depth, and $D_{\text{mult}}$ is its multiplicative depth. The weights $w_{\text{ops}}$, $w_{\text{depth}}$, and $w_{\text{mult}}$ reflect the importance of each metric in the cost function. For our experiments, we set all weights to 1. \edit{
We study sensitivity to these
weights in Sec.~\ref{ablation:reward_weights} and show that this configuration enables the RL agent to generate code that has minimal execution time.
}
The components of the cost function are defined below.

\shrink[0.5]
\paragraph{Operations Cost ($C_{\text{ops}}$):} This term quantifies the computational cost of all of the operators used in the expression. We assign a numerical cost to each operator. These numerical costs reflect its relative performance compared to other operators in the BFV scheme. The relative order of these operations is guided by former empirical studies on the cost of FHE operations~\cite{FHEBench}. The costs are structured to incentivize vectorization:
1) \emph{Vector Additions/Subtractions:} These have the lowest cost (1), as they represent the most efficient parallel operations.
2) \emph{Vector Multiplications:} These have a higher cost than vector additions/subtractions, and usually cost $100\times$ the cost of vector additions, and therefore we assign a cost of (100) to them.
3) \emph{Rotations ($<<$):} Rotations typically have a cost that is $0.5 \times$ to $1 \times$ of the cost of a vector multiplication (depending on the parameters). We assign a cost of (50) to rotations, reflecting that rotations might be cheaper than multiplications but always more expensive than additions. This incentivizes rotations over multiplications, since they have a cost that is usually less than or equal to the cost of multiplications.
4) \emph{Scalar Operations (+,-,*):} These are assigned a high cost (250) to reflect their inefficiency. While in this particular case, the value 250 is not derived empirically, we chose this high cost for scalar operations to incentivize the RL to generate vectorized code.

The total \textbf{$C_{\text{ops}}$} is the sum of the costs of all the nodes in the expression tree.

\shrink[0.5]
\paragraph{Circuit Depth ($D_{\text{circuit}}$)}
A greater depth implies more sequential operations are performed on a ciphertext, which leads to a larger accumulation of noise. Minimizing depth is therefore critical, as it directly impacts the noise.

\shrink[0.5]
\paragraph{Multiplicative Depth ($D_{\text{mult}}$):}
This metric is critical as noise growth in the BFV scheme is directly proportional to the multiplicative depth. Therefore, to minimize the noise, we aim to minimize the multiplicative depth.

\edit{
The three terms have different numerical ranges: $C_{\text{ops}}$ can be in the
hundreds, while $D_{\text{circuit}}$ and $D_{\text{mult}}$ are typically in the
single digits. As a result, even with equal weights the objective tends to
prioritize reducing operation cost (runtime). Increasing $w_{\text{depth}}$ and
$w_{\text{mult}}$ shifts the policy toward circuits with lower noise growth, but this often comes at the cost of additional operations, leading to slower execution. We quantify this
speed/noise trade-off in Sec.~\ref{ablation:reward_weights}.
}

\shrink[0.5]
\subsubsection{Reward Structure}\label{reward_structure}
To guide the agent's learning process, the reward signal is composed of two parts. The first is an immediate reward provided after every action during training, which guides local, incremental improvements. The second is a terminal reward provided once the entire optimization sequence for a given expression has ended (end of episode), and which guides the agent toward a globally optimal solution.

   \emph{i) Step Reward ($R_{\text{step}}$):} After each action, the agent receives an immediate reward calculated as the relative percentage improvement in the cost function. If the cost of the expression before the action is $C_t$ and after is $C_{t+1}$, the reward of the step is:

   {
   \shrink[2.5]
   \CodeSize 
   \begin{equation*}
   R_{\text{step}} = \frac{C_t - C_{t+1}}{C_t}
   \end{equation*}
   }
    
   \emph{ii) Terminal Reward ($R_{\text{final}}$):}
    The reward is given at the end of an episode (when the agent selects the \texttt{END} action, or the pre-defined maximum steps limit for the episode is reached). This reward is based on the total percentage reduction in cost from the initial state of the expression ($C_{\text{initial}}$) to its final state ($C_{\text{final}}$):

    \vspace{0.4cm}
    {
    \shrink[2]
    \CodeSize 
    \begin{equation*}
    R_{\text{final}} = \left( \frac{C_{\text{initial}} - C_{\text{final}}}{C_{\text{initial}}} \right) \times 100
    \end{equation*}
    }

    \edit{
    \paragraph{Terminal reward motivation.}
    We initially trained the CHEHAB RL agent using only the step reward and found that the
    policy often became locally greedy: it favored rewrite sequences that yield small
    early improvements and avoided transformations that temporarily increase cost but
    unlock larger gains later. This behavior is expected because PPO optimizes
    discounted returns, which biases learning toward immediate improvements.
    To better handle delayed credit in rewrite optimization, we add a terminal reward
    based solely on the final optimized expression, which encourages sequences that
    optimize end-to-end circuit quality rather than only short-horizon gains.
}

    \edit{
    We quantify the impact of removing $R_{\text{final}}$ (i.e., using only the immediate reward) in Sec. \ref{ablation:reward_no_final}, showing that the combined \textit{immediate$+$terminal} reward yields better end-to-end execution time compared to step-only reward.}

    

\subsection{Actor-Critic Networks \label{sec:actor-critic}}

In this work, we adopt the actor-critic RL architecture. To manage the complexity of the action space, we learn a \emph{hierarchical stochastic policy} (actor), which decomposes the action \(a\) into two components: a rewrite rule \(r\) and an application location \(p\). This policy is defined by two conditional probability distributions: \( \pi_{\theta_1}(r \mid s) \) for rule selection and \( \pi_{\theta_2}(p \mid s, r) \) for location selection. Alongside the policy, we learn a value function (critic) \( V_{\phi}(s) \), which estimates the expected return from a given state \(s\). The policy and critic networks are implemented as separate deep neural networks.

\shrink[0.5]
\subsubsection{Rule Selection Network} The first network of the actor is the rule selection network. It is a Multi-Layer Perceptron (MLP) that takes, as input, the 256-dimensional state embedding generated by the Transformer encoder (Section~\ref{state}). The network consists of two fully connected hidden layers with 128 and 64 neurons, respectively, each using the ReLU activation function. The network's final output layer is followed by a softmax, which produces a probability for each available action from the action space. After invalid actions (non-matching rules) are masked, the agent samples from this resulting probability distribution to select its next action.

\shrink[0.5]
\subsubsection{Location Selection Network}

Once the rule network selects a rewriting rule, the Location Network selects the location within the IR for its application. The network selects whether the rule is applied to its 1st match in the IR, 2nd match, 3rd match, etc.
It is implemented as an MLP that takes two inputs: the 256-dimensional state embedding and the output from the actor network representing the chosen rule. Its architecture consists of two fully connected hidden layers, each with 64 neurons, followed by ReLU activation. The final output layer produces a probability distribution over locations,
from which the final position is sampled.

\begin{figure}[t]
    \shrink
    \centering
    \includegraphics[width=\linewidth]{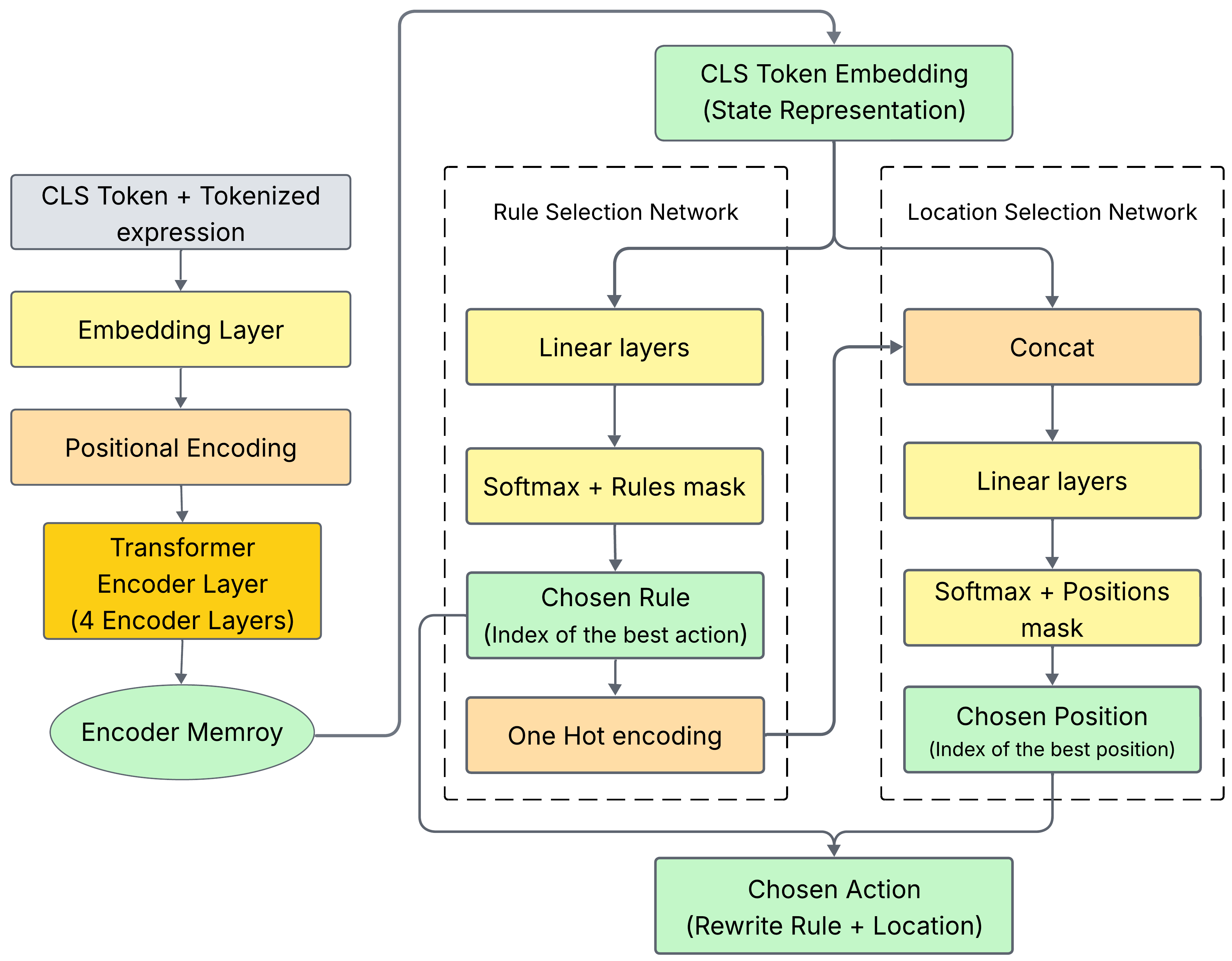}
    \caption{Architecture of the Policy Network}
    \Description{A diagram showing the neural network architecture used for the policy.}
    \label{fig:policy}
\end{figure}

\shrink[0.5]
\subsubsection{The Critic (Value Network)} The critic network estimates the value function, $V(s)$, which predicts the expected cumulative future reward from the current state. It is an MLP that takes the state embedding as input. Its architecture consists of 3 fully-connected hidden layers with 256, 128, and 64 neurons, respectively, each with a ReLU. The final layer is a linear output layer that produces a single scalar value representing the value estimate, $V(s)$. This estimate is used during training to assess the actor's actions and guide the learning.

%% file: 5-training.tex
\shrink[1]
\section{Training Data\label{data}}

\shrink[0.5]
A key challenge for training an effective \edit{CHEHAB} RL agent is \edit{the absence of datasets for FHE programs. Consequently, we need to generate} a dataset of expressions that are both diverse and contain realistic code patterns that appear in practice.

\shrink[0.5]
\paragraph{Limitations of Random Data Generation}
Training on purely random expressions is not ideal. Randomly generated expressions do not follow the same distribution of programs encountered in practice. They often lack the specific computational patterns that appear in practice (e.g., common sub-expressions, opportunities for factorization, and structured arithmetic that can be vectorized).

\shrink[1]
\paragraph{LLM-guided Generation}
To address the previous limitation, we use \emph{Large Language Models (LLMs)} to generate the training data, \edit{This design choice is important in practice: Sec.~\ref{ablation:llm_vs_random}
shows that training on LLM-generated programs yields substantially better policies
than training on randomly generated programs.\ifshowextra The full prompt template and a link to the generated training dataset are provided in Appendix~\ref{app:prompt}.\fi} We guide the LLM to synthesize a dataset that is diverse and rich in optimizable code patterns. Since the LLM is trained on a large corpus of realistic code, it tends to generate code that follows the same distribution of realistic code. For our data generation, we used the \emph{Gemini 2.5 Flash} model. The prompt provided to the LLM is structured to give it sufficient context about the problem domain, and includes:

    \emph{i) Syntax and Semantics:} A formal description and multiple examples of the CHEHAB IR, covering scalar operations, vector constructors (\texttt{Vec}), and vector operations (e.g., \texttt{VecAdd}).
    
    \emph{ii) Rewrite Rule Examples:} Examples of the rewrite rules from our TRS are included to guide the LLM on the specific patterns we intend to optimize.
    
    \emph{iii) Real World Examples:} To ground the synthesis process in practical computations, the prompt contains real-world kernels as examples. Examples provided in the prompt include:
    a) {Union Cardinality:} Computes the cardinality of the bitwise OR of two 4-bit vectors;
    b) {Squared Difference:} Computes the element-wise squared error between two vectors;
    c) {4x4 Matrix Addition:} A common linear algebra kernel that performs element-wise addition between two matrices. The provided examples do not include any of the benchmarks used in the evaluation.
    
    \emph{iv) High-Level Goal:} A description of the optimization objective and an explicit request for structural diversity in the generated expressions.

\edit{
\shrink[2.5]
\paragraph{Synthesizing CHEHAB IR Directly.}
We also considered generating CHEHAB DSL programs (embedded in C++) and then lowering them
through the compiler to CHEHAB IR. In practice, directly generating CHEHAB IR was more
reliable and easier to validate: the IR has a small, closed vocabulary and
simple typing constraints, making it easier for the LLM to produce syntactically
valid programs. In contrast, generating CHEHAB DSL (which is embedded in C++) requires emitting compilable code
(including boilerplate C++ code), and validating each sample requires
invoking a C++ toolchain and running DSL-to-IR lowering, which significantly
slows the generation loop. Direct IR generation also simplifies deduplication
and benchmark exclusion, since we can directly parse and canonicalize each program before adding it to the corpus.
}

\shrink[1]
\paragraph{Post Processing}
Expressions generated by the LLM are filtered to ensure their validity and uniqueness:

    \emph{i) Parsing and Validation:} Expressions that cannot be parsed by our IR parser (due to syntactic errors) are discarded.
    
    \emph{ii) Uniqueness Filtering:} To ensure diversity and avoid redundancy, we implement a uniqueness filter based on the canonical representation of expressions. This canonical representation is obtained using our \Tokenization{} (\Token{}) tokenization.
    Generated expressions may be semantically equivalent but use different variable names (e.g., {\CodeSize \texttt{(Vec (+ x (* y z)))}} and {\CodeSize \texttt{(Vec (+ a (* b c)))}}). After \Token{} tokenization, they become equivalent.
    A newly generated expression is added to the dataset only if its canonical form was not generated before.
    
    \emph{iii) Exclusion of Benchmarks:} To avoid training the RL agent on programs identical to those found in the benchmark, we remove any generated expression that is identical to one of the benchmarks. \edit{This ensures the evaluation measures generalization to unseen programs.}
    To do this, we compute the canonical form of each program in our benchmark suite (Sec.~\ref{s:evaluation}) using \Token{} tokenization. Any expression in the generated dataset that matches a benchmark is removed.

Our approach ensures that the dataset is not only syntactically correct but also semantically relevant and rich in the patterns our agent needs to learn to optimize.
We generated \emph{15,855} unique expressions using the LLM and used this dataset to train the RL agent.

\edit{
\shrink[1]
\paragraph{Generalizability and benchmark separation.} We strictly separate training and evaluation. The LLM-generated training dataset contains 15,855 unique expressions. Each expression is parsed and converted to its \Token{} canonical form, which normalizes away differences in identifiers and constants and collapses common syntactic variants. We use this canonical form to
deduplicate the dataset and to remove any generated expression whose canonical form matches a benchmark expression. This ensures that no benchmark program (or a trivially rewritten variant) appears in the training data, so evaluation reflects generalization to unseen programs.}






%% file: 6-evaluation.tex
\shrink[1]
\section{Evaluation}\label{s:evaluation}

We evaluate our proposed approach in three ways:

\shrink[0.5]
\paragraph{Quality of Programs Generated}
We compare code generated by CHEHAB RL with that produced by the Coyote~\cite{Malik2023Coyote} compiler.
The comparison focuses on the execution time of the generated circuit, the number of homomorphic operations, the depth, the multiplicative depth, and noise (noise accumulated by the benchmark, which we want to minimize).\ifshowextra The definition of the noise metric and details about how we measure it are in Appendix~\ref{noise}.\fi
We limit our evaluation to FHE compilers that support unstructured code.
Porcupine~\cite{Cowan2021Porcupine} is excluded as it is not publicly available (we reached out to its authors but could not obtain a copy).

\shrink[0.5]
\paragraph{Compilation Efficiency} We report the end-to-end compilation time for CHEHAB RL and Coyote.

\shrink[0.5]
\paragraph{Ablation Study} We perform an ablation study to evaluate the following: LLM-generated vs random data; 2) CHEHAB RL vs CHEHAB; 3) \Token{} vs BPE tokenization; 4) Flat vs hierarchical action space; 5) GRU vs Transformer encoder.

\begin{figure*}[t]
    \centering
    \includegraphics[width=\textwidth]{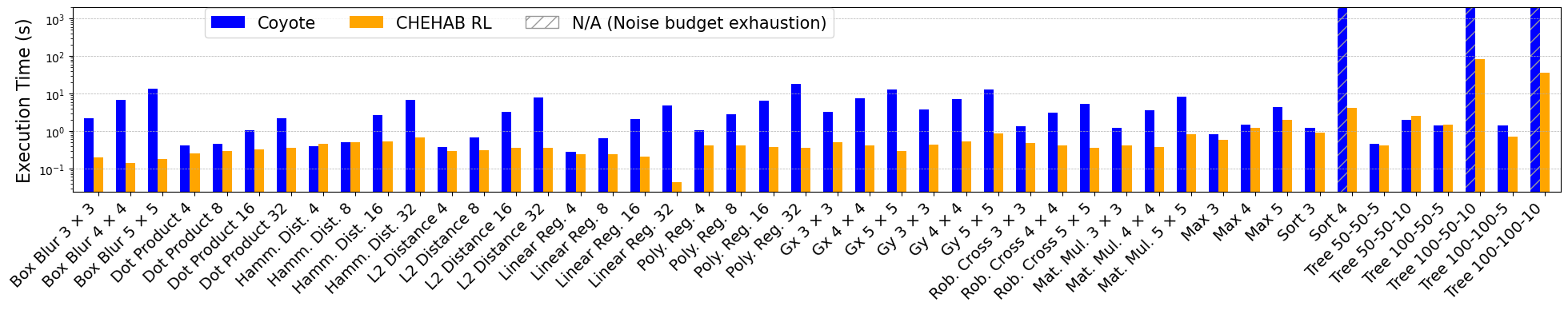}
    \shrink[3.5]
    \caption{Semi-log plot of execution time of the benchmarks, comparing code generated using our compiler and Coyote.}
    \shrink[1.2]
    \Description{A semi-logarithmic plot comparing the execution time of various benchmarks. It contrasts the performance of code generated by the CHEHAB compiler against the Coyote compiler, with the y-axis showing time on a log scale.}
    \label{fig:kernel}
\end{figure*}

\begin{figure*}[t]
    \centering
    \includegraphics[width=\textwidth]{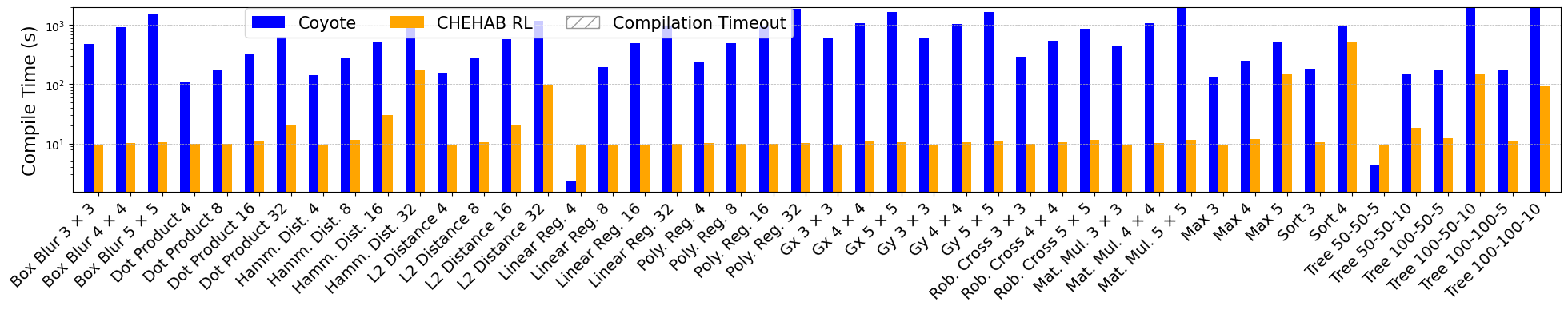}
    \shrink[3.5]
    \caption{Semi-log plot of compilation time of the benchmarks using our compiler and Coyote.}
    \shrink[0.4]

    \label{fig:compil_time_plot}
    \Description{A semi-logarithmic bar chart comparing the compilation times of various benchmarks. It shows the time taken by the CHEHAB compiler versus the Coyote compiler.}
\end{figure*}
\begin{figure*}[t]
    \centering
    \includegraphics[width=\textwidth]{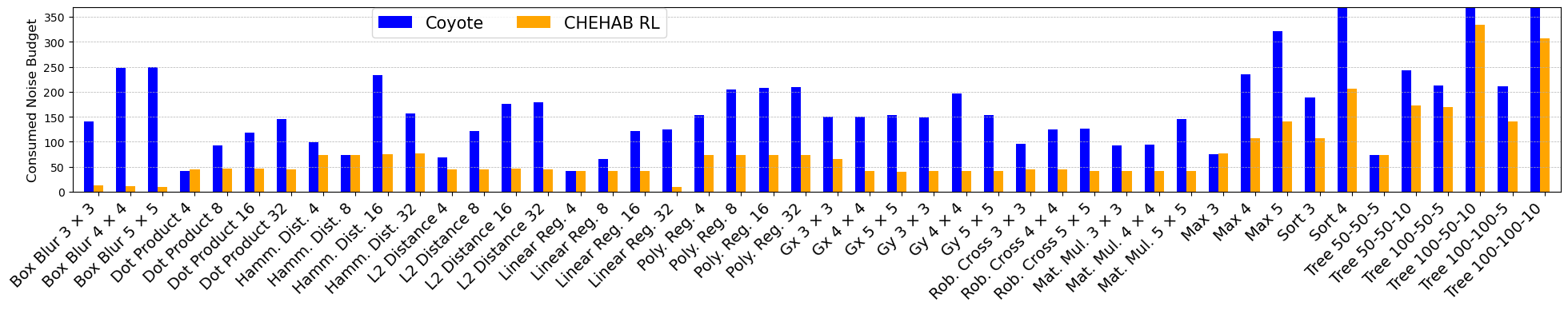}
    \shrink[3]
    \caption{Semi-log plots comparing the Consumed Noise Budget in Coyote and CHEHAB RL.}
    \shrink[0.4]
    \label{fig:noise_consume}
    \Description{A semi-logarithmic plot comparing the Consumed Noise Budget (CNB) usage between the Coyote compiler and CHEHAB RL across various benchmarks. The y-axis represents the noise budget consumed.}
\end{figure*}


\subsection{RL Agent Training} \label{sec:agent_training}
We use the Proximal Policy Optimization (PPO) algorithm to train the agent \cite{Schulman_PPO_2017} used by CHEHAB RL. We train it for 2 million timesteps (43 hours), on a single multi-core CPU.
More details about \ifshowextra the training and \fi the hyperparameters in
\ifshowextra Appendix~\ref{training} and \fi
Table~\ref{tab:ppo_hyperparameters}.

\subsection{Benchmarks}

We evaluate CHEHAB RL on three benchmark suites. For each benchmark in the benchmark suites, we vary the input size to evaluate the scalability of the compiler. The larger the input is, the larger the program is, since FHE code is fully unrolled. The three benchmark suites are:

\paragraph{Porcupine Benchmark Suite} This is a set of kernels commonly used in linear algebra, machine learning, and image processing. They were used to evaluate Porcupine~\cite{Cowan2021Porcupine}. They include image processing filters: Box Blur, the x and y gradients of an image (\emph{Gx} and \emph{Gy}), and Robert Cross (\emph{Rob. Cross.}). The Porcupine benchmark suite also includes Dot Product, Hamming Distance (\emph{Hamm. Dist.}), L2 Distance, Linear Regression (\emph{Linear Reg.}), and Polynomial Regression (\emph{Poly. Reg.}), which are building blocks for ML applications. We include all of the benchmarks that were used to evaluate Porcupine.

\paragraph{Coyote Benchmark Suite} It includes kernels that were used to evaluate Coyote~\cite{Malik2023Coyote}. These kernels are: 1) matrix multiplication (\emph{Mat. Mul.}); 2) a sorting algorithm (\emph{Sort}) that sorts a list of elements; This sorting algorithm is an unstructured code that implements sorting using a tree (as described by Malik et al.~\cite{Malik2023Coyote}).
3) \emph{Max} kernel. This is an unstructured code that finds the maximum element in a list.
Coyote also includes three other kernels, which are the dot product, the L2 distance, and a convolution, but these are common with the Porcupine benchmark suite, and therefore, we do not evaluate them again.
All of the benchmarks that were used to evaluate Coyote are included in our benchmark suite.

\shrink[0.25]
\paragraph{Randomly Generated Irregular Polynomials} The Coyote compiler also uses a set of randomly generated unstructured polynomials for evaluation. We use the same methodology for evaluation. This set of polynomials is a stress test to further investigate the ability of the compiler to vectorize code in the absence of a regular structure in the code. Several polynomials are generated as random arithmetic expression trees. In these polynomials, a polynomial is named tree-X-Y-Z, where high values for X-Y indicate that the polynomial tree is full, complete, and well-balanced, making it easier to vectorize, while low values for X-Y indicate that the tree is sparse and imbalanced, making it harder to vectorize. Z indicates the depth of the tree of the polynomial.\ifshowextra We provide more details about these polynomials in Appendix~\ref{irregularPolynomials}.\fi

\shrink[1]
\subsection{Input Layout Transformation Before Encryption \label{sec:data-layout-optimization}}

\edit{Vectorization often requires rearranging the layout of the input data so that operands line up in the same ciphertext slots. If performed after encryption, this layout transformation translates into extra ciphertext rotations (and masking), which are expensive. CHEHAB therefore moves this step to the client: inputs are permuted/packed in plaintext and only then encrypted, so the server-side computation starts from ciphertexts already in the desired layout and avoids the corresponding homomorphic rotations. Unless stated otherwise, the evaluation reports results with this optimization enabled, and we provide results without this optimization in the appendix.}

\shrink[1]
\subsection{Environment Setup}

We evaluate our benchmarks on a 2.40 GHz Intel(R) Xeon(R) CPU E5-2680 v4 with 256 GB DDR4 of memory running CentOS Linux 8. The encryption parameters for CHEHAB and Coyote are set as follows: n = 16384 (polynomial modulus degree) and q is the default coefficient modulus proposed by the SEAL library for a standard 128-bit security level.
We disable the CHEHAB pass for automatic rotation key selection\ifshowextra (presented in Appendix~\ref{appendix:RotationKeysSelection})\fi, since Coyote does not have a similar pass.
We also disable the use of the blocking technique \cite{Malik2023Coyote} for both compilers, CHEHAB and Coyote, since both of them perform blocking in the same way. Blocking is a technique that involves vectorizing smaller kernels separately and then composing the resulting vectorized programs. While this technique enables the two compilers to scale to larger programs, our goal is to compare the fundamental algorithms proposed by each compiler on a single block. Since the blocking method is identical in the two compilers, comparing on a single block will better illustrate the fundamental differences between the two approaches. 
\edit{By default, CHEHAB transforms the data layout of input data before
encryption (as described in Sec.~\ref{sec:data-layout-optimization}). Results when the data layout of input data is transformed after encryption are presented
in the appendix in Table~\ref{tab:kernel}.}
We also use the same SEAL library in both compilers (SEAL version 4.1). We run all the experiments 30 times and report the median execution times and compilation times. The compilation timeout for both compilers was 7200 seconds.

\shrink[1]
\subsection{Evaluation on the Benchmarks}

We compare CHEHAB RL with Coyote on our benchmark suite, measuring the execution time of optimized code (Fig.~\ref{fig:kernel}), compilation time (Fig.~\ref{fig:compil_time_plot}), and consumed noise budget (Fig.~\ref{fig:noise_consume}). Fig.~\ref{fig:kernel} shows that CHEHAB RL produces faster code than Coyote for most benchmarks. On average, CHEHAB RL generates code that is $\ExecutionTimeSpeedupOverCoyote{}\times$ faster in execution (geometric mean) than Coyote.
For example, on \texttt{Poly. Reg. 32}, code generated by CHEHAB RL is $50\times$ faster than code generated by Coyote, and faster by $114\times$ on \texttt{Linear Reg. 32}. 
This speedup is due to having fewer operations in code generated by CHEHAB RL, for example, in \texttt{Poly. Reg. 32}, CHEHAB RL generated a circuit with just 3 additions and 2 ct-ct multiplications (ciphertext-ciphertext multiplications). In contrast, Coyote generates a much larger circuit with 12 additions, 2 ct-ct multiplications, 1 subtraction, 173 ct-pt multiplications (ciphertext-plaintext multiplications), and 134 rotations. Coyote appears to generate a complex data layout that requires extensive rotations and ct-pt multiplications to execute, where CHEHAB RL finds a simpler, more direct vectorization. This pattern holds for \texttt{Linear Reg. 32} as well, where the agent produces a circuit with 2 additions, 1 ct-ct multiplication, and 1 ct-pt multiplication, while Coyote generates one with 4 additions, 1 ct-ct multiplication, 3 subtractions, 46 ct-pt multiplications, and 44 rotations.
Image-processing kernels also show substantial gains: code generated by CHEHAB RL for \texttt{Gx 5x5} is 42$\times$ faster than Coyote.
In this benchmark, Coyote over-rotates data, leading to a higher number of operations and a slower circuit.
For \texttt{Tree 50-50-10}, Coyote generates a circuit that runs faster; this is due to the high number of expensive ct-ct multiplications that CHEHAB RL generates. The generated circuits that Coyote generates, despite having a higher number of total operations (21 adds, 19 plaintext multiplications, 7 rotations), require only 7 ct-ct multiplications. In contrast, the circuit produced by CHEHAB RL, while more compact (18 adds, 2 rotations), requires 15 ct-ct multiplications.

Fig.~\ref{fig:compil_time_plot} shows that despite producing circuits that are faster to execute, the compilation time for CHEHAB RL is also consistently faster than Coyote's.
On average, the CHEHAB RL compilation process is $\CompilationTimeSpeedupOverCoyote{}\times$ faster (geometric mean) than Coyote. A notable exception to this trend is the \texttt{Tree 50-50-5} and \texttt{Linear Reg. 4} benchmarks, where Coyote's compilation time is faster than CHEHAB RL. These benchmarks are small, so Coyote's search algorithm can rapidly explore the small search space. In contrast, our RL agent requires a series of steps to apply its learned policy. The overhead of invoking the neural network for each of these sequential steps results in a longer compilation time for these small benchmarks. However, as demonstrated across the rest of the benchmark suite, this overhead for the RL agent is outweighed by its higher scalability to larger circuits.

Fig.~\ref{fig:noise_consume} compares the noise budget consumed by the circuits generated by CHEHAB RL and Coyote.
The results show that code generated by CHEHAB RL consistently consumes less noise budget compared to Coyote.
On average, code generated by CHEHAB RL consumes $\NoiseReductionOverCoyote{}\times$ less noise budget (geometric mean) compared to Coyote.
For example, code generated by CHEHAB RL for \texttt{Poly. Reg. 32} consumes only 73 bits from the available noise budget (which is 369 bits), leaving a remaining budget of 296 bits, while Coyote's much larger and complex circuit consumes 210 bits. In different benchmarks such as \texttt{Sort 4} and two of the \texttt{Polynomial Tree} benchmarks, the circuit generated by Coyote exhausts the entire noise budget and fails to execute, while CHEHAB RL successfully produces valid, runnable circuits for all benchmarks, leaving a safe remaining noise budget in each case.

A more detailed comparison, including operation counts for all benchmarks, is available in Table~\ref{tab:kernel} in Appendix~\ref{appendix:benchmarkeval}.
\begin{figure*}[!t]
    \centering
    \includegraphics[width=\textwidth]{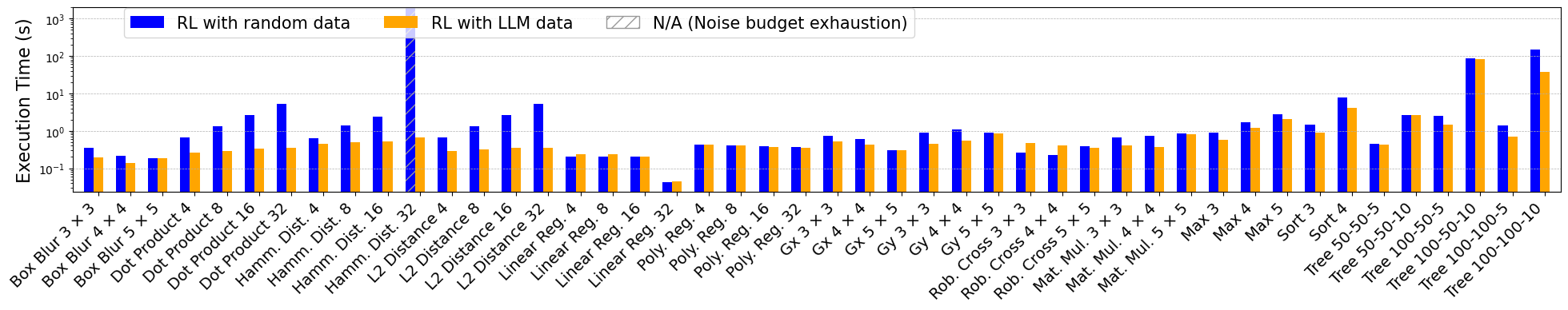}
    \shrink[3.5]
    \caption{Semi-log plots comparing the execution times when using an RL agent trained with randomly generated data and an RL agent trained with data generated by an LLM}
        \shrink[1.2]

    \label{fig:random_data_plot}
    \Description{A semi-logarithmic plot comparing the execution times of two Reinforcement Learning agents. One agent was trained using randomly generated data, while the other was trained using data generated by a Large Language Model (LLM).}
\end{figure*}

\shrink[1]
\subsection{Ablation Study}
\begin{figure*}[!t]
    \centering
    \includegraphics[width=\textwidth]{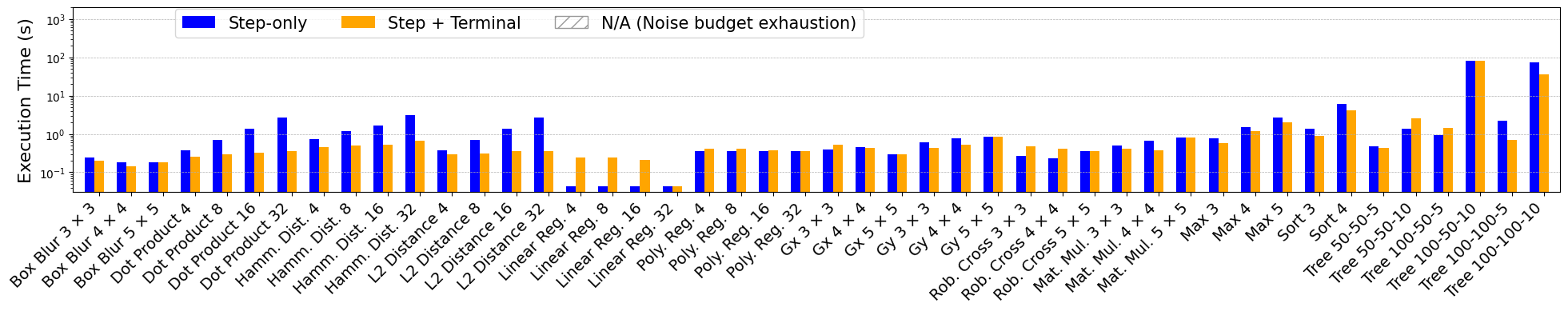}
    \shrink[3.5]
    \Description{A semi-logarithmic plot comparing the execution time performance of two reinforcement learning agents. One agent is trained using only step-based rewards, while the other (the authors' approach) uses a combination of step and terminal rewards.}
    \caption{\edit{Semi-log plot of benchmark execution time, comparing an agent trained with
step-only reward against our step$+$terminal reward.}}
    \label{fig:final_vs_nofinal_reward}
    \shrink[1.2]
\end{figure*}

\edit{
\shrink[2.5]
\paragraph{Step vs. step$+$terminal reward.}\label{ablation:reward_no_final}
Our reward signal is composed of an immediate step reward and a terminal reward (Sec. \ref{reward_structure}),
where the step reward provides local feedback after each rewrite and the terminal reward provides a global signal at the end of the optimization episode.
We ablate this design by training an agent with only the immediate reward (step wise) and comparing it to our default agent using immediate $+$ terminal rewards.
Using both components is beneficial: \textit{immediate$+$terminal} achieves a $1.291\times$ better execution time (geometric mean) than using the immediate reward alone.
This shows that the terminal reward is necessary to align the policy with end-to-end circuit quality, whereas purely local feedback can over-emphasize short-horizon improvements that do not translate into the best final circuit. Figure~\ref{fig:final_vs_nofinal_reward} shows the per-benchmark execution time.
}

\edit{
\shrink[2.5]
\paragraph{Reward weight sensitivity.}\label{ablation:reward_weights}
We further ablate the reward design by varying the weights of the cost function in Sec. \ref{reward_cost},
i.e., $(w_{\text{ops}}, w_{\text{depth}}, w_{\text{mult}})$.
We compare the default $(1,1,1)$ against a set of new weights: $(1,50,50)$, $(1,100,100)$ and $(1,150,150)$.
We include configurations with large $w_{\text{depth}}$ and $w_{\text{mult}}$ because operation costs in $C_{\text{ops}}$ are already numerically large, and thus depth-based penalties must be scaled to have a comparable influence when explicitly biasing the policy toward lower-depth circuits.
While the new weight variants have a slightly lower noise consumption (they consume $0.911$-$0.941\times$ the noise of $(1,1,1)$),
our default weight configuration, $(1,1,1)$, yields significantly better runtime: it is $1.396$-$1.487\times$ faster in execution time (geometric mean) than the other variants. This is summarized in Table~\ref{tab:reward_weight_sensitivity}.
}

\begin{table}[t]
\centering
\scriptsize
\resizebox{\columnwidth}{!}{%
\begin{tabular}{lcc}
\toprule
\edit{$(w_{\text{ops}}, w_{\text{depth}}, w_{\text{mult}})$}
& \edit{Exec. time ($\times$ vs. $(1,1,1)$)}
& \edit{Noise ($\times$ vs. $(1,1,1)$)} \\
\midrule
\edit{$(1,50,50)$}      & \edit{1.426$\times$} & \edit{0.941$\times$} \\
\edit{$(1,100,100)$}    & \edit{1.487$\times$} & \edit{0.935$\times$} \\
\edit{$(1,150,150)$}    & \edit{1.396$\times$} & \edit{0.911$\times$} \\
\bottomrule
\end{tabular}%
}
\caption{\edit{Reward weight sensitivity.}}
\label{tab:reward_weight_sensitivity}
\end{table}

\shrink[0.5]
\paragraph{LLM-generated vs. Random Data}\label{ablation:llm_vs_random}
To quantify the effectiveness of using an LLM to generate data, we compare the performance of our RL agent (trained on LLM-generated data) against the same agent trained on a dataset of randomly generated data. The random code generator uses a standard approach to generate code with a uniform distribution \ifshowextra(described in Appendix~\ref {randomdata})\fi.
Fig.~\ref{fig:random_data_plot} compares the two approaches. The agent trained on LLM-generated data produces FHE circuits that are faster. For instance, on the \texttt{L2 Distance 32} benchmark, CHEHAB RL trained on LLM-generated data produces code that is 13$\times$ faster compared to CHEHAB RL trained on randomly generated data. Similar performance gaps of an order of magnitude are observed across the larger instances of \emph{Hamm. Dist.} and \emph{Dot Product}.
This difference in performance is due to the quality of the learned policies. The agent trained on LLM-generated data learns to generate more efficient circuits with fewer expensive homomorphic operations. For example, in the \texttt{L2 Distance 32} case, the circuit generated by the LLM-trained agent contains only 5 additions and 1 multiplication, compared to 28 additions and 28 multiplications in the circuit generated by the agent trained on random data. This demonstrates that the LLM-generated data helps the RL agent learn a policy that performs better on realistic programs, since it provides expressions with relevant, optimizable structures, that more closely mirror real program distributions.


\shrink[0.5]
\paragraph{CHEHAB RL vs CHEHAB}
To ensure that the improvements in reducing latency in CHEHAB RL are due to the RL TRS, and not to other optimizations in the original CHEHAB, we compare the execution times of code generated by the original CHEHAB and CHEHAB RL. Fig.~\ref{fig:execution_time_chehab_vs_rl} (Appendix) shows the results. As we see, in most cases, code generated by CHEHAB RL is faster. In some instances, most notably \texttt{Gx 3x3}, the original CHEHAB is better than CHEHAB RL due to a sub-optimal decision made by the learned policy. An examination of the generated circuit shows that the RL agent chose to apply one rotation to align data for the vectorization of two ct-pt multiplications. While the CHEHAB RL successfully vectorized the operations, the computational cost of the required rotation was greater than the performance benefit gained from vectorizing the two ct-pt multiplications.


\shrink[0.5]
\paragraph{\Token{} vs. BPE Tokenization}
We also evaluate the use of the \Tokenization{} (\Token{}) tokenization for tokenizing the inputs of the embedding model. This experiment compares our agent, which uses \Token{} tokenization, against the same agent that uses a standard Byte-Pair Encoding (BPE) tokenizer. The BPE tokenizer was trained on a corpus of 5 million randomly generated IR expressions to build its vocabulary \ifshowextra(details about random IR code generation are in Appendix~\ref{randomdata})\fi. Once the BPE tokenizer is trained, we train the two agents using our dataset of LLM-generated data. Fig.~\ref{fig:training_time_bpe_vs_dynamic} in the appendix shows that the RL agent that uses \Token{} tokenization finishes its 2 million steps training in 43 hours, compared to the RL agent trained with BPE, which takes 68 hours,
showing that our tokenization helps accelerate the training.

\shrink[0.5]
\ifshowextra
\paragraph{Flat vs. Hierarchical Action Spaces.}\label{ablation:flat_vs_heir}

We also compared the hierarchical action space against a flat one that enumerates rule-location pairs (i.e, each rule is duplicated multiple times, once for each location). The learning curves in Fig.\ref{fig:flat_vs_hier} in the appendix show consistently higher rewards and quicker learning for the hierarchical agent.
\fi
\ifshowextra
\shrink[1]
\paragraph{GRU vs Transformer}

To validate our choice of a Transformer encoder for the state representation, we compared its encoding capabilities against a baseline recurrent architecture, the Gated Recurrent Unit (GRU). Our experiments show that the transformer architecture was able to learn a better representation with less training compute. Details about this experiment are in Appendix~\ref{appendix:ablation_encoder}.
\fi

%% file: 7-related_work.tex
\section{Related Work\label{sec:realted}}

This section compares CHEHAB RL and state-of-the-art compilers.
Table~\ref{tab:related} summarizes the differences, which we discuss in detail in this section.
\ifshowextra 
In addition, we discuss the use of RL in compilers, tokenization methods, and random data generation methods in Appendix~\ref{appendix:related}.
\fi

\paragraph{FHE Compilers for Automatic Vectorization}

Recent work \cite{Viand2022HECOFH, Malik2023Coyote, Cowan2021Porcupine, Dathathri2019CHET} focuses on the problem of transforming FHE programs that use scalar variables into vectorized programs (batching).
Our work is similar to these compilers since it also takes a scalar code and vectorizes it. Unlike HECO~\cite{Viand2022HECOFH}, CHET~\cite{Dathathri2019CHET}, ANT-ACE~\cite{li2025antace}, Orion~\cite{ebel2025orion}, HEIR~\cite{heir_compiler} and Qiwu \cite{zhongcheng2025qiwu} which only support the vectorization of structured (loop-based) code, CHEHAB RL is designed to support both structured and unstructured code.
Compared to Coyote~\cite{Malik2023Coyote}, our approach scales better.
Unlike Porcupine, in which the user has to manually provide the data layout of the vectorized code, our approach computes the best data layout automatically.

\paragraph{FHE Compilers for Circuit Optimization}

These compilers do not vectorize their input code. They focus on applying other optimizations (other than vectorization). These compilers include EVA~\cite{Dathathri2020}, Ramparts~\cite{Archer2019} and HECATE~\cite{Lee2022}.
The main challenges that they target are automatic parameter selection, the scheduling of ciphertext-maintenance operations, reducing the multiplicative depth of circuits, and reducing the number of operations in a circuit. 
Our approach is complementary to the above compilers. First, it supports code vectorization, and therefore, code vectorized by our approach can be passed to these compilers for further optimization. Our approach also offers the ability to reduce the depth of circuits, which they do not address.
\ifshowextra 
Appendix~\ref{appendix:related} provides a more detailed comparison with these compilers.
\fi

\input{tables/related_work}
\paragraph{General Purpose Vectorization in non-FHE Programs}

Superword-Level Parallelism is a classical technique used for code vectorization~\cite{slp}.
It processes a sequence of scalar instructions to create vector packs or groups of isomorphic instructions that can be packed together into vectors.
Since it does not depend on the presence of data-parallel loops in the code, it is well-suited for vectorizing unstructured code.
However, while creating vector packs, SLP does not consider the high cost of rotations in FHE programs, which results in code that has a high number of rotations, making it impractical for the domain of FHE, where rotations are expensive.
goSLP~\cite{goSLP} is a state-of-the-art SLP approach that formulates the vectorization problem as an Integer Linear Programming (ILP) problem. However, it operates at a level closely tied to the target architecture, where the vectors are restricted to a maximum width of four elements. One of its key limitations is that while it efficiently handles the pairwise packing of statements, extending this to pack more than two statements makes the problem intractable for current solvers. In contrast, our approach for FHE supports significantly larger vectors (in the thousands), which is necessary for the domain of FHE since vectors in FHE are significantly larger.
VeGen \cite{veGen} extends SLP by introducing lane-level parallelism, tracking how individual lanes execute computations. This enables VeGen to account for rotation costs when constructing vector packs. However, this reasoning is local, as VeGen does not consider the impact of instruction packing on subsequent rotations.
Other work~\cite{10.1145/996893.996853,10.1145/3297858.3304059,10.1145/1133255.1133996} also addresses code vectorization but is not designed for the field of FHE and does not assume a high cost for rotations. In addition, in this work, our goal is to develop an FHE compiler that not only vectorizes code but also reduces the instruction latency and noise growth while being more scalable.

\paragraph{RL for Code Optimization}

RL has been explored for compiler code optimization~\cite{mlcompsurvey,trofin2021mlgo,VenkataKeerthy_2023,pecenin2019optimization,10.5555/3314872.3314896,haj2020neurovectorizer,brauckmann2021polygym,cummins2022compilergym}. Systems such as \emph{Halide RL}~\cite{pecenin2019optimization} train agents to select schedules for image processing code, while \emph{Tiramisu RL}~\cite{10.5555/3314872.3314896} applies polyhedral transformations to computational kernels using an RL-based policy. More specialized approaches include \emph{NeuroVectorizer}~\cite{haj2020neurovectorizer}, which learns to select the optimal vectorization and interleaving factors for loops on SIMD architectures using learned code embeddings, and \emph{Polygym}~\cite{brauckmann2021polygym}, which frames affine loop transformation in the polyhedral model as a Markov Decision Process. Frameworks such as \emph{CompilerGym}~\cite{cummins2022compilergym} have provided RL environments for tasks such as LLVM phase ordering.
While previous work addresses traditional code optimizations, it does not consider the unique constraints of Fully Homomorphic Encryption. Our work is the first to formulate FHE optimization as an RL problem, where the agent must learn a policy that navigates the trade-offs between vectorization, cryptographic noise accumulation, and the cost of operations. 
\ifshowextra 
Appendix~\ref{appendix:related} provides more details on using RL for code optimization.
\fi

\paragraph{Automatic Code Optimization in Compilers}

Automatic code optimization for loop nests has been widely explored.
Examples of state-of-the-art methods include the use of the polyhedral model and deep-learning-based methods~\cite{Iri88,thies_unified_2001,bondhugula_practical_2008,baghdadi2015PhD,baghdadi2018tiramisu1,polly,chen2018learning,zheng2020ansor,merouani2024looper,hakimi2023hybrid,mlgo}. Such work does not address FHE code optimization, though.

\ifshowextra\else
\section{Additional Details and Experiments}
Additional details, experiments, and ablation studies are available in the arXiv version of this paper: \url{https://arxiv.org/}.
\fi 

%% file: tables/related_work.tex
\begin{table}[t]
    \fontsizefortables
    \setlength\tabcolsep{2pt}
    \caption{\textsc{Comparison with related work.}}
    \Description{A table comparing CHEHAB RL with related compilers (HECO, Porcupine, Coyote, Ramparts, EVA, HECATE) across features like vectorization, multiplication depth reduction, and supported encryption schemes.}
    \shrink[2]
    \begin{tabular}{p{2.7cm}|c|c|c|c|c|c|c}
        \hline
        
        \textbf{Feature} & \textbf{\rotatebox{90}{\ CHEHAB RL\ }} & \textbf{\rotatebox{90}{\ HECO\ }} & \textbf{\rotatebox{90}{\ Porcupine\ }} & \textbf{\rotatebox{90}{\ Coyote\ }} & \textbf{\rotatebox{90}{\ Ramparts\ }} & \textbf{\rotatebox{90}{\ EVA\ }} & \textbf{\rotatebox{90}{\ HECATE\ }}\\\hline

        \textbf{Vec. struct. code} & \yes  & \yes & \yes & \yes & \no & \no & \no \\\hline

        \textbf{Vec. unstruct. code} & \yes  & \no & \yes & \yes & \no & \no & \no \\\hline

        \textbf{Reduce mul. depth} & \yes & \yes & \yes & \no & \yes & \no & \no \\\hline

        \textbf{Auto datalayout} & \yes  & \yes & \no & \yes & \yes & \yes & \yes \\\hline

        \textbf{Data driven} & \yes  & \no & \no & \no & \no & \no & \no \\\hline 

        \textbf{Scheme} & BFV  &  BFV  & BFV & BFV & BFV & CKKS & CKKS \\
        \textbf{  }     &      &  CKKS &     &     &     &      &      \\
        \textbf{ }      &      &  BGV  &     &     &     &      &     \\\hline

    \end{tabular}
    \label{tab:related}
\end{table}

%% file: 8-conclusion.tex
\section{Conclusion}

This paper introduces a novel framework that leverages RL to automate FHE code optimization. Our proposed approach trains an RL agent to learn a policy for applying a sequence of rewriting rules to automatically vectorize scalar FHE code while reducing instruction latency and noise growth. We show that our approach generates code that is $\ExecutionTimeSpeedupOverCoyote{}\times$ faster than Coyote, accumulates $\NoiseReductionOverCoyote{}\times$ less noise, while it takes $\CompilationTimeSpeedupOverCoyote{}\times$ less time in compiling code, enabling better scalability.

\shrink[1]
\section*{Acknowledgment}

This research was partly supported by the Center for Cyber Security (CCS) at New York University Abu Dhabi.
It was also partly supported by the Center for Artificial Intelligence and Robotics (CAIR) at New York University Abu Dhabi, funded by Tamkeen under the NYUAD Research Institute Award CG010.
It was also partly supported by the Federation of Arab Scientific Research Council under contract number \mbox{ARICA23\_787}.
The research was carried out on the High-Performance Computing resources at New York University Abu Dhabi.

%% file: 9-appendix.tex
\ifshowextra 

\section{More Detailed Background}

\subsection{\edit{Term Rewriting System (TRS)}}\label{TRSbackground}

\edit{A term rewriting system (TRS) is a set of rewrite rules that transform an expression to a new form (rewrite it into a new expression).}
\edit{Let us take a simple example of a TRS system to illustrate how it works. Let us assume that we want to simplify arithmetic expressions and let us assume that we have the following rule set} $S = \{t-t \rightarrow \, 0,\; t + 0 \rightarrow \, t\}$.
\edit{Let us use this rule set \(S\) to rewrite and simplify the expression $b + (a - a)$.}
\edit{The rule $t-t \rightarrow 0$ can be applied to simplify the expression $a - a$. After applying the rule, we get the new expression: $b + 0$. Now we can apply the rule $t + 0 \rightarrow t$ to simplify the previous expression into $b$.}

\subsection{Fully Homomorphic Encryption\label{appendix:fhe}}

Fully Homomorphic Encryption (FHE) enables computations directly on encrypted data \cite{armknecht2015guide, e3}.  
A \textit{homomorphism} is a function between two groups that preserves their structure \cite{dummit2004abstract}.  
In the context of FHE, these groups correspond to elements in the plaintext space $\mathcal{P}$ and the ciphertext space $\mathcal{C}$.
Let us define the encryption and decryption functions as $E(\cdot): \mathcal{P} \mapsto \mathcal{C}$ and $D(\cdot): \mathcal{C} \mapsto \mathcal{P}$, respectively.  
Given an operation $\circ$ on plaintexts and its homomorphic counterpart $\circledcirc$ on ciphertexts, for plaintexts $m_a, m_b \in \mathcal{P}$ with encryptions $c_a = E(m_a), c_b = E(m_b) \in \mathcal{C}$, the homomorphic property ensures that operations on encrypted data remain consistent with their plaintext counterparts, i.e., $m_a \circ m_b = D(c_a \circledcirc c_b)$.

\subsubsection{BFV Scheme\label{bfvapp}}
The BFV encryption scheme \cite{bfv_a:2012/144} is an FHE scheme based on the Ring-Learning With Errors (RLWE) problem \cite{rlwe}.
In BFV, the plaintext space is defined as $\mathcal{P} = R_t = \mathbb{Z}_t[x]/(x^n+1)$, while the ciphertext space is $\mathcal{C} = R_q \times R_q$, where $R_q = \mathbb{Z}_q[x]/(x^n+1)$.  
Here, $n$ is the polynomial modulus degree, and $t$, $q$, and $x^n+1$ denote the plaintext, ciphertext, and polynomial moduli.
Typical values for $t$ range from 16 to 32 bits, while $q$ varies between 100 and 900 bits.
The degree $n$ is usually a power of two, commonly between $2^{10}$ and $2^{16}$, as recommended by the \emph{Homomorphic Encryption Standard} \cite{albrecht2018}.

\subsubsection{FHE Batching\label{appendix:batching}}

Batching enables evaluation of a function on \(n\) blocks of data \cite{Brakerski2012} at once by encoding a vector of \(n\) messages in a single plaintext polynomial in \(R_t\). Under the assumption \(t \equiv 1 \mod 2n\), \(x^n + 1\) factors into linear polynomials modulo \(t\), i.e. \(x^n + 1 \equiv \prod_{i=1}^{n} (x - a_i) \mod t\), with \(a_i = a^{2i-1}\) being the $2n^\text{th}$ primitive root of unity modulo \(t\), and \(t = \prod_{i=1}^{n} t_i\) where \(t_i\) are prime ideals with basis \((t, x-a_i)\). Using CRT on ideals, we have the following isomorphism: 
\[
R_t = R/(t) \overset{CRT}{\cong}R/t_1 \times \dots \times R/t_n
\]
Thus, evaluating a function once over \(R/(t)\) evaluates the same function on smaller plaintext spaces \(R/t_1,\dots,R/t_n\).

\fi

\ifshowextra 

\section{Selection of Rotation Keys}
\label{appendix:RotationKeysSelection}

\subsection{Importance of Selecting the Rotation Keys}
Galois keys (a.k.a. rotation keys) are essential for performing rotation operations in FHE.
Each distinct rotation step requires its own key—for example, rotating a ciphertext by $s_1$ and another by $s_2$ (where $s_1 \neq s_2$) necessitates two separate rotation keys. 
When a program involves many unique rotation steps, generating and transmitting all keys becomes costly, as each key is several megabytes in size.
A common approach to mitigate this is to generate a subset of rotation keys and express other rotations as combinations of these.
For instance, generating only the rotation key for step $s = 1$ allows any rotation to be performed as repeated single-step rotations.
This reduces key generation and communication costs but significantly increases execution time and noise for large rotation steps.
A better trade-off is to balance the costs of key generation and communication, as well as the cost of executing rotations during homomorphic operations.

\subsection{Method for Selecting Rotation Keys}
\label{subsec:rotation_keys_selection}

After code optimizations, CHEHAB generates an IR containing the necessary rotations for which keys must be generated.
Previous work~\cite{Cowan2021Porcupine, Viand2022HECOFH, Archer2019, Lee2022} does not address automatic rotation key selection, relying instead on the FHE library's default, which generates $2 \log_2(n)$ keys.
This approach can be suboptimal, as some applications may require fewer than \(2 \log_2(n)\) rotation keys.
Instead, CHEHAB selects the rotation keys to be generated and ensures that the number of keys does not exceed a user-defined upper bound $\beta$, which defaults to \(2 \log_2(n)\).

Let \(\chi\) be the set of all rotation steps used in the program. 
Our goal is to decompose the steps in \(\chi\).
We use the non-adjacent form (NAF) representation of each step \(s\) in \(\chi\) to decompose it. For example, possible decompositions of a step $s = 3$ are obtained by calculating $NAF(3)$. The NAF representation writes s as a sum of powers of two, where each coefficient is either +1 or -1, and no two nonzero digits are adjacent. For example, NAF(3) = 4 - 1; NAF(5) = 4 + 1.

Once we calculate the NAF representation for each step s, we collect the decompositions of the steps. For each rotation step $s$, let $\Gamma_s$ denote the set of decompositions of s obtained from its NAF (e.g., for $s = 3$ we get $\Gamma_3 = \{-1, 4\}$).

We then select a subset $\Omega$ of the rotation steps in \(\chi\) that will be decomposed via their NAF. The remaining rotation steps that are not decomposed form another set, denoted $\chi_f$.

The final set of rotation keys to be generated is the union of the keys for the non-decomposed rotations and those derived from the NAF decompositions.

Consider the following example:

{
\CodeSize
\setlength{\jot}{\baselineskip}
\setlength{\jot}{4pt} 
\shrink[2]
\begin{gather*}
\chi = \{1,2,3,4,5,6,7,9,10,12,11,13,15\} \\
n = 16, \quad 2 \times \log_2(n) = 8, \quad \beta = 9
\end{gather*}
\shrink[2]
}

The NAF decompositions of steps in $\chi$ are as follows:

{
\CodeSize
\setlength{\jot}{\baselineskip}
\setlength{\jot}{4pt} 
\shrink[1]
\begin{gather*}
NAF(1) = 1, \quad NAF(2) = 2, \quad NAF(3) = -1 + 4 \\
NAF(4) = 4, \quad NAF(5) = 1 + 4, \quad NAF(6) = -2 + 8 \\
NAF(7) = -1 + 8, \quad NAF(9) = 1 + 8, \quad NAF(10) = 2 + 8 \\
NAF(12) = -4 + 16, \quad NAF(11) = -1 - 4 + 16 \\
NAF(13) = 1 - 4 + 16, \quad NAF(15) = -1 + 16 \\
\Gamma_1 = \{1\}, \quad \Gamma_2 = \{2\}, \quad \Gamma_3 = \{-1, 4\} \\
\Gamma_4 = \{4\}, \quad \Gamma_5 = \{1, 4\}, \quad \Gamma_6 = \{-2, 8\} \\
\Gamma_7 = \{-1, 8\}, \quad \Gamma_9 = \{1, 8\}, \quad \Gamma_{10} = \{2, 8\} \\
\Gamma_{12} = \{-4\}, \quad \Gamma_{11} = \{-1, -4\} \\
\Gamma_{13} = \{1, -4\}, \quad \Gamma_{15} = \{-1\}
\end{gather*}
\setlength{\jot}{\baselineskip} 
\shrink
}

A valid set of rotation steps selected for decomposition is:
{
\shrink
\CodeSize
\[
\Omega = \{1,2,3,4,5,6,7,9,12,15\}
\]
}

In that case, the final state is:
{
\CodeSize
\setlength{\jot}{4pt} 
\begin{gather*}
\chi_f = \{10, 11, 13\} \\
\Gamma_{tot} = \bigcup_{s \in \Omega} \Gamma_s= \{1,2,4,-1,-4,8\}
\end{gather*}
\setlength{\jot}{\baselineskip} 
\shrink 
}

\noindent where $\Gamma_{tot}$ is the set of decompositions of steps in \(\chi\) selected for decomposition.
At the end, we end up with a smaller number of rotation keys to generate, since we just need to generate 9 keys (a key for each step in \(\chi_f \cup \Gamma_{tot}\)), instead of generating 13 keys (a key for each rotation step in \(\chi\)).

\fi

\section{\edit{CHEHAB DSL syntax and semantics.}}\label{sec:chehab_dsl}
\edit{
CHEHAB is an \emph{embedded} DSL implemented via C++ operator overloading on \texttt{Ciphertext} and \texttt{Plaintext}.
Although CHEHAB programs are written in C++, only a restricted subset of C++ expressions and helper functions constitutes the DSL.
We therefore summarize this \emph{practical DSL subset} as a compact grammar for readability, while keeping the full operator list in Table~\ref{tab:operations} as a reference.
}
\subsection{\edit{Core expression grammar (pseudo-grammar).}}
\edit{
Programs in CHEHAB construct expression graphs over \texttt{Ciphertext} or \texttt{Plaintext} values:}
\[
\begin{aligned}
e ::= &\; x \mid k \mid (e) \mid e \ \oplus\ e \mid -e \mid e \ \rho\ k \\
     &\mid f(e) \mid g(\{e_1,\ldots,e_n\})
      \mid e.\texttt{set\_output}(\texttt{"name"})
\end{aligned}
\]
\edit{where $x$ is an identifier, $k$ is an integer literal, $\oplus \in \{+, -, *\}$, and
$\rho \in \{\texttt{<<}, \texttt{>>}\}$.}

\edit{The supported helper functions are:}
\[
\begin{aligned}
f &\in \{\texttt{square},\ \texttt{reduce\_add},\ \texttt{reduce\_mul},\ \texttt{SumVec},\ \texttt{encrypt}\},\\
g &\in \{\texttt{add\_many},\ \texttt{mul\_many}\}.
\end{aligned}
\]

\edit{
\paragraph{Pseudo-semantics.}
Expressions are typed as \texttt{Ciphertext} or as \texttt{Plaintext}.
Binary operators denote the corresponding homomorphic operations (ct-ct or ct-pt variants depending on operand types),
unary \texttt{-} denotes negation, and \texttt{<<}/\texttt{>>} denote slot rotations by a constant offset.
Helper functions expand into compositions of these primitives (e.g., \texttt{square} and \texttt{exponentiate} as repeated
multiplication, \texttt{add\_many} and \texttt{mul\_many} as reductions over a set of expressions, while
\texttt{reduce\_add}, \texttt{reduce\_mul} and \texttt{SumVec} as structured reductions).
Integer literals are permitted in expressions via implicit \texttt{Plaintext} construction (and \texttt{encrypt} when needed).
Finally, \texttt{set\_output("name")} marks an expression as a program output for compilation.
}
\subsection{List of Operations in the CHEHAB DSL}

Table~\ref{tab:operations} provides the full list of operations in the CHEHAB DSL as well as their signatures and descriptions. 

\setlength\tabcolsep{3pt}
\setlength\arrayrulewidth{0.3pt}
\begin{table}[h!t]
    \centering
    \fontsizefortables
    \caption{Operations of the CHEHAB domain-specific language (ct: ciphertext, pt: plaintext, int: integer).}
    \shrink
    \label{tab:operations}
    \begin{tabular}{ccp{5.3cm}}
        \toprule
        \textbf{Op.} & \textbf{Signature} & \textbf{Description} \\
        \midrule
        \multirow{3}{*}{+} & ct $\times$ ct $\rightarrow$ ct & \multirow{3}{=}{Element-wise addition} \\
        & ct $\times$ pt $\rightarrow$ ct \\
        & pt $\times$ ct $\rightarrow$ ct \\
        \midrule
        
        \multirow{1}{*}{+} & ct $\times$ int $\rightarrow$ ct & \multirow{1}{=}{Add int value to all elements of ct} \\
        \midrule

        \multirow{3}{*}{-} & ct $\times$ ct $\rightarrow$ ct & \multirow{3}{=}{Element-wise subtraction} \\
        & ct $\times$ pt $\rightarrow$ ct \\
        & pt $\times$ ct $\rightarrow$ ct \\
        \midrule

        \multirow{1}{*}{-} & ct $\times$ int $\rightarrow$ ct & Subtract int value from all elements of ct \\
        \midrule

        \multirow{1}*{-} & \multirow{1}*{ct $\rightarrow$ ct} & Negation of each element of the argument \\
        \midrule

        \multirow{2}*{<{}<} & \multirow{2}*{ct $\times$ int $\rightarrow$ ct} & Rotation of ciphertext to \\
        & & the left with the given step \\
        \midrule

        \multirow{2}*{>{}>} & \multirow{2}*{ct $\times$ int $\rightarrow$ ct} & Rotation  of the ciphertext to  \\
        & & the right with the given step  \\
        \midrule
        
        \multirow{3}{*}{*} & ct $\times$ ct $\rightarrow$ ct & \multirow{3}{=}{Element-wise multiplication} \\
        & ct $\times$ pt $\rightarrow$ ct \\
        & pt $\times$ ct $\rightarrow$ ct \\
        \midrule

        \multirow{1}{*}{*} & ct $\times$ int $\rightarrow$ ct & Multiply all ciphertext elements by an int
        \\
        
        \bottomrule
    \end{tabular}
    \shrink
\end{table}
\setlength\tabcolsep{3pt}

\ifshowextra
\section{Code Generation\label{appendix:codegen}}

To reduce memory consumption, CHEHAB minimizes the creation of temporary ciphertext and plaintext objects using the primitive \emph{inplace} provided by the SEAL API. This primitive enforces the reuse of memory occupied by dead objects, similar to how a compound assignment operator works.
To compile a program, the user runs the DSL code. When the DSL code is run, it creates the IR AST, runs the compiler passes on the IR, and generates optimized code at the end (C++ code). This code is further compiled using a C++ compiler. The generated binary code is the final outcome of the compilation process and can be run like any other binary. This approach is common in implementing DSLs embedded in C++.

\fi

\ifshowextra
\section{Rewriting Rules \label{rules}}
The transformations available to the RL agent are defined as a set of rewriting rules.
These rules include rules for vectorization, algebraic simplification, and rotation.

    \paragraph{Vectorization Rules for Isomorphic Subexpressions.} These rules pack scalar arithmetic operations into single vector instructions. They search for isomorphic element‑wise scalar expressions and rewrite them as a single vector instruction. For example, a vectorization rule for addition is:
    
    {\CodeSize
    \begin{center}
    \texttt{(Vec (+ a b) (+ c d))} \(\Rightarrow\) \texttt{(VecAdd (Vec a c) (Vec b d))}
    \end{center}
    }
    This rewriting rule replaces two scalar additions (left-hand side) with one vector addition operating on newly constructed vectors of the operands (right-hand side). Similar rules exist for multiplication, subtraction, and negation.
    
    \paragraph{Vectorization Rules for Non-isomorphic Subexpressions.}
    While powerful, the previous vectorization rules require isomorphic subexpressions. To handle common patterns of non-isomorphic subexpressions, our set of rules also includes general rules for vectorizing such non-isomorphic patterns.
    For each arithmetic operation \emph{op}, we have a rule that matches whenever an expression contains a mix of operations (non-isomorphic subexpressions), provided that the operation \emph{op} appears more than once. The rule then vectorizes all instances of the operation \emph{op}. It also moves any non-matching sub-expression (containing operators other than \emph{op}) into the first operand vector. It then pads the second operand vector with the appropriate identity element (e.g., 1 for multiplication, 0 for addition).
    For example:
    
    {\CodeSize
    \begin{center}
    \texttt{(Vec (* a b) (* c d) (- f g))} \(\Rightarrow\) \\
    \texttt{(VecMul (Vec a c (- f g)) (Vec b d 1))}
    \end{center}
    }
    Here, the non-isomorphic vectorization rule for multiplication matches because there are two * operations. It packs the two multiplications into a single vector multiplication operation, and leaves \texttt{(- f g)} in the 1st operand, and pads the second operand vector with 1, the multiplicative identity.
    
    Since the RL agent is trained to predict rewriting rules that maximize the global reward, it automatically weights the cost and benefit of applying this type of vectorization and selects this class of rules only if they are beneficial.
    
    \paragraph{Simplification and Algebraic Rules:} This category includes standard algebraic rules that simplify expressions, reduce computational complexity, or transform expressions into a different form (that will be useful in later simplifications). These rules aim to lower the circuit's depth and minimize the number of operations. Examples include:

    \begin{itemize}
        \item \emph{Arithmetic Simplification:} These rules simplify arithmetic expressions, replacing complex or multiple operations with simpler, equivalent ones. Examples of these rules include factorization 
        
        {\CodeSize
        \begin{center}
            $\texttt{(+ (* x y) (* x z))} \;\Rightarrow\; \texttt{(* x (+ y z))}$,
        \end{center}
        }
        
        \noindent identity elimination
        
        {\CodeSize
        \begin{center}
            $\texttt{(x * 1)} \Rightarrow \texttt{x}$,
        \end{center}
        }   
            
        \noindent absorption rules
        
        {\CodeSize
        \begin{center}
            $\texttt{(x * 0)} \Rightarrow \texttt{0}$,
        \end{center}
        }

        plaintext consolidation

        {\CodeSize
        \begin{center}
            $\texttt{(* (pt a) (* (pt b) x))} \;\Rightarrow\; \texttt{(* (pt (a*b)) x)}$,
        \end{center}
        }
        
        \noindent where a and b are plaintexts.
    
        \item \emph{Arithmetic Transformations:} These rules transform arithmetic expressions in a way that enables their simplification later. Examples include commutativity, associativity, and distribution rules.

        \item \emph{Circuit Balancing:} 
        These rules balance expression trees, reducing their depth (and noise accumulation). For example, a left-leaning tree of multiplications can be balanced to reduce the multiplicative depth:

        {\CodeSize
        \begin{center}
            \texttt{(VecMul x (VecMul y (VecMul z t)))} \(\Rightarrow\) \\
            \texttt{(VecMul (VecMul x y) (VecMul z t))}
        \end{center}
        }
    \end{itemize}
    
    \paragraph{Rotation Rules:} Data alignment is critical in FHE, and rotations are the primary mechanism for moving data within a packed ciphertext. A naive approach to vectorizing unstructured code would require the RL agent to discover a long and potentially inefficient sequence of low-level rotation, masking, and arithmetic operations to correctly align data.
    To address this, our set includes rules that transform high-level computational patterns directly into efficient, \textit{composite dataflow structures}. These rules encapsulate what would otherwise be multiple low-level steps into a single, high-level transformation.
    For example, we consider this expression:
    
    {\CodeSize
        \begin{center}
            \texttt{(Vec ( (+ (* a b) (* c d)) (+ (* e f) (* g h)) ) )}
        \end{center}
    }
        
    A standard vectorization strategy, without leveraging rotations, would result in the following structure:
    
    {\CodeSize
        \begin{center}
            \texttt{(VecAdd  (VecMul (Vec a e ) (Vec b f) ) \\ (VecMul ( Vec c g ) (Vec d h ) ) )}
        \end{center}
    }
        
    This optimized expression requires two vector multiplication operations and one vector addition operation.
    However, a more sophisticated strategy, enabled by our rotation-based rules, can find a more efficient solution:
    
    {\CodeSize
        \begin{center}
            \texttt{(VecAdd V (<{}< V 2))}\\
            where V:
            \texttt{(VecMul (Vec a c e g) (Vec b d f h))}
        \end{center}
    }
    
    The final result can then be computed by adding this vector V to a rotated version of itself, which effectively sums the required pairs of products into the first two slots.
    This alternative strategy requires only one vector multiplication operation, one vector addition, and one rotation. Since a vector rotation is cheaper than a vector multiplication in FHE, this second approach is better. By including such composite transformation rules, they help the RL agent to discover these globally optimal strategies that a simpler vectorizer would miss.

\subsection{How did we design our rewriting rules?}

To construct the rules for the TRS, we began by collecting the rules from Halide's TRS~\cite{10.1145/3150211,Newcomb2020}. Halide is an industrial compiler used for the optimization of image processing and deep learning pipelines. Halide’s expression space includes operations that are not natively supported by fully homomorphic encryption (FHE), such as comparison, division, and modulo operations. Additionally, performance considerations and optimization goals in Halide differ from ours, as it operates on plaintext data. Therefore, we selected only the rules that are compatible with FHE. Next, we expanded this initial ruleset manually and developed new rules aimed at reducing the number of operations, rotations, depth, and multiplicative depth in FHE programs.

\fi

\ifshowextra
\section{\edit{LLM Prompt Template for CHEHAB IR Synthesis}}
\label{app:prompt}

\edit{The full prompt template used to synthesize CHEHAB IR expressions is presented below. The template encodes (i) CHEHAB IR syntax and validity checks, (ii) explicit constraints on vector width and expression depth, (iii) a structural-diversity requirement beyond alpha-renaming, and (iv) worked examples and rewrite-rule context to bias generation toward expressions that benefit from rewrite-based optimization.}

\begin{lstlisting}[style=promptstyle, title=\edit{Synthesis Prompt}]
[SYSTEM ROLE]
You are a rigorous validator for CHEHAB IR expressions. First ANALYZE then GENERATE. Enforce structural uniqueness beyond variable renaming.

[GENERATION PROTOCOL]
######  CHEHAB IR Generation Protocol ######

You will output **5 structurally unique** `(Vec ...)` expressions for RL training.
Every expression **must** honour all rules below. If any check fails, discard the draft and regenerate before replying.

1. Core Vector Form
--------------------------------------------------------------
- Start with `(Vec` and contain **exactly {vec_size} sub-expressions**.  
  Skeleton -> `(Vec expr0 expr1 ... expr_n)`.  
- No nested `(Vec ...)` inside a sub-expression.  
- Sub-expression depth: **4 <= depth <= 20**.  

2. Syntax / Operator Rules
--------------------------------------------------------------
- Balanced parentheses.  
- Operators: `+` and `*` are strictly binary; `-` may be unary or binary.  
- Variables match `[a-z][0-9]_*[0-9]*` (e.g. `in_1_0`).  
- **No numeric literal 0** anywhere.
- No degenerate single-term parentheses such as `(x)`.

3. Semantic / Structural Rules
--------------------------------------------------------------
- **Structural uniqueness** after canonicalising w.r.t. history/examples.  
- **Operation asymmetry**: avoid identical operator trees across siblings.  
- Expressions must not be trivially vectorisable; a sequence of rewrite rules 
  (see Sec 7) should improve depth or multiplicative depth.

4. Mandatory Generation Checklist
--------------------------------------------------------------
1. Draft 5 candidate `(Vec ...)` lines.  
2. Count elements == {vec_size}.  
3. Validate parentheses & operator set.  
4. Compute depths & variable counts.  
5. Canonicalise and check for structural duplicates.  
6. Output the clean expressions--one per line, no commentary.
7. At least 3 expressions must have depth > 10. 

5. Worked Example Breakdown (Real World Motifs)
--------------------------------------------------------------
Below are full CHEHAB IR programs illustrating real computations. 
*Do NOT copy or trivially rename them.*

- Union-Cardinality (size 1):
(Vec (+ (+ (+ ( - ( + v1_0 v2_0 ) ( * v1_0 v2_0 ) ) ... )))

- Squared Difference (size 4):
(Vec ( * ( - v1_0 v2_0 ) ( - v1_0 v2_0 ) ) ... )

6. Rewrite Rules Context
--------------------------------------------------------------
[The LLM is provided with the following rules to bias generation toward 
optimizable patterns:]
Rewrite { name: "add-vectorize-2", searcher: (Vec (+ ?a0 ?b0) (+ ?a1 ?b1)), applier: (VecAdd (Vec ?a0 ?a1) (Vec ?b0 ?b1)) }
Rewrite { name: "mul-vectorize-4", searcher: (Vec (* ?a0 ?b0) (* ?a1 ?b1)), ... }
Rewrite { name: "comm-factor-1", searcher: (+ (* ?a ?b) (* ?a ?c)), applier: (* ?a (+ ?b ?c)) }

... [84 total rules provided] ...

7.  Final Output
--------------------------------------------------------------

- Produce **exactly 5** valid, structurally distinct (Vec ...) expressions.
- At least 3 of the generate expressions **Must have a depth > 6**.
- The Generated expressions where they match vectorization rules , or rules that enables vectorization.
- The Generated expressions must match real world computations and programs or similair **not fully random computations**.
- One expression **per line**, raw text—no numbering, comments, or styling.
- Range them from moderately simple to deeply nested.
- If any candidate breaks §§1–3, discard & regenerate before responding."""

\end{lstlisting}

\edit{
\paragraph{Prompt refinement.}
We refined the prompt iteratively by manually reviewing generated programs and adding constraints to reduce invalid outputs and duplicates. Using this protocol, we generated a training dataset of 15,855 unique expressions, which is available at: 
}\textbf{\url{https://raw.githubusercontent.com/Modern-Compilers-Lab/CHEHAB/refs/heads/main/RL/fhe_rl/datasets/final_llm_dataset.txt}}.
\fi

\ifshowextra
\section{Training Details\label{training}}
The agent was trained on the LLM-generated dataset.
We used a single node with a 2.40 GHz Intel(R) Xeon(R) CPU E5-2680 v4 (14 cores) and 256 GB DDR4 of memory running CentOS Linux 8 for the training. We used the \emph{Stable-Baselines3 library}~\cite{stable-baselines3} to implement our RL training.
To accelerate the collection of experience, we leverage parallel environments. We run multiple independent instances of the optimization environment in parallel (8 in our case), allowing the agent to collect a diverse batch of trajectories simultaneously.
To implement this, we leverage multi-processing where each environment runs in its own process, and multiple episodes are collected in parallel across processes. This speeds up the training process.

Each training episode corresponds to the optimization of a single expression and is limited to a maximum length of 75 steps. We found this limit to be sufficient, as the optimization sequences for the majority of expressions in our training and evaluation sets finished well before reaching 75 steps. The agent was trained for a total of 2 million timesteps, and the training took 43 hours.
\fi

\begin{table}[htbp]
\fontsizefortables
\centering
\caption{PPO Hyperparameters for Training.}
\label{tab:ppo_hyperparameters}
\begin{tabular}{ll}
\hline
\textbf{Hyperparameter} & \textbf{Value} \\ \hline
Learning Rate & $1 \times 10^{-4}$ \\
Discount Factor ($\gamma$) & 0.99 \\
GAE Lambda ($\lambda$) & 0.95 \\
PPO Clip Range ($\epsilon$) & 0.2 \\
Update Epochs & 20 \\
Steps per Update & 2048 \\
Batch Size & 256 \\
Number of Environments & 8 \\ \hline
\end{tabular}
\end{table}

\ifshowextra
\section{More Details about the Evaluation}
\label{appendix:evaluation}

\subsection{Noise Measurement \label{noise}}

A freshly encrypted ciphertext begins with a large noise budget, and each homomorphic operation consumes a part of it. When the budget reaches zero, decryption fails. Our goal is to minimize the noise budget consumed by code.

To measure the noise budget consumed in each benchmark, we use the \texttt{Decryptor::invariant\_noise\_budget} function provided by the Microsoft SEAL library. In the BFV scheme, every ciphertext carries an error term (noise) that grows as homomorphic operations are performed. SEAL reports the remaining noise budget (in bits).

In our setup, using a polynomial modulus degree of 16384, a freshly encrypted ciphertext has an initial noise budget of 369 bits. After running each 
benchmark, we query the remaining noise budget and measure the difference from the initial value. The difference is the consumed noise that we report (in bits).

We use a 20-bit plaintext modulus and select the coefficient modulus using SEAL’s helper

\begin{center}
    \texttt{CoeffModulus::BFVDefault(poly\_modulus\_degree)}
\end{center}

This returns a coefficient modulus with a total size of 389 bits (\texttt{total\_coeff\_modulus\_bits = 389}), which is consistent with a theoretical maximum initial noise budget of \(389 - 20 = 369\) bits (i.e., \texttt{total\_coeff\_modulus\_bits - plain\_modulus\_bits}).

\subsection{Random Code Generation\label{randomdata}}

Our random code generator recursively constructs IR expressions. We control this construction by two parameters: maximum depth and vector size. The generator builds expressions by sampling a mixture of scalar operations, vector operations, rotations, and vector constructors. Sampling is balanced across all combinations of depth (1–15) and vector size (1–32), so the model encounters a wide range of shapes and packing patterns.

Concretely, the generator works as follows:
\begin{enumerate}
  \item It samples a target vector size \(n\) and a depth budget \(d\).
  \item Starting from a root, it grows an expression tree until the tree depth budget is reached.
  \item At each node, it picks an operator from \{\emph{scalar op}, \emph{vector op}, \texttt{Vec} constructor\} that is type-compatible and respects the chosen vector size \(n\).
  \item Leaves are instantiated with variables or integer constants.
  \item The resulting expression is parsed and type-checked into the compiler IR; invalid samples are discarded.
\end{enumerate}

\subsection{Randomly Generated Unstructured Kernels\label{irregularPolynomials}}

We borrow the following description of the randomly generated irregular polynomials from the original Coyote~\cite{Malik2023Coyote} paper, which proposed the benchmark.
Several polynomials are randomly generated to evaluate as arbitrary arithmetic expression trees. The trees are generated according to three different regimes to cover different kinds of programs:

\begin{itemize}
    \item Dense, homogeneous: The expression tree is both full and complete, and all the operations are isomorphic. In principle, this represents a best case for vectorization. We refer to these as tree-100-100.
    \item Dense, nonhomogeneous: The expression tree is both full and complete, each operation has a 50/50 chance of being an add or a multiply. Hence, while the trees are structurally similar, the heterogeneity of operations means that vectorization opportunities are restricted. We refer to these as tree-100-50.
    \item Sparse: Many operations have one leaf node input; the tree is not very balanced. In principle, this represents a worst case for vectorization, where Coyote must work hard to find vectorizable computation. We refer to these as tree-00-50.    
\end{itemize}

The last number in the names of the polynomials indicates the depth of the tree. For example, tree-100-100-5 indicates a dense, homogeneous tree with a depth equal to 5.
\fi
\section{Benchmark Evaluation Results\label{appendix:benchmarkeval}}

\input{tables/notation}

Table~\ref{tab:kernel} presents several metrics regarding the depth and number of operations for three cases: 1) the initial (naive) implementation of the benchmarks 2) CHEHAB RL's generated code for each one of the benchmarks, and 3) Coyote's generated code. The table compares the three cases in terms of circuit depth ($\cup$), multiplicative depth ($\cup^\otimes$), number of ciphertext-ciphertext multiplications ($\otimes$), number of rotations ($\circlearrowright$), number of ciphertext-plaintext multiplications ($\odot$), and number of ciphertext additions ($\oplus$). We also report CHEHAB's and Coyote's compilation times CT (s) and their Consumed Noise budget (CN), where lower values are better. The notation used in Table~\ref{tab:kernel} is presented in Table \ref{tab:notation}. The most important columns in the table are the multiplicative depth ($\cup^\otimes$), the number of rotations ($\circlearrowright$), the compilation time (CT) for the two compilers, and the Consumed Noise budget (CN). We highlight them in bold.
\begin{figure}[t]
    \centering
    \includegraphics[width=\linewidth]{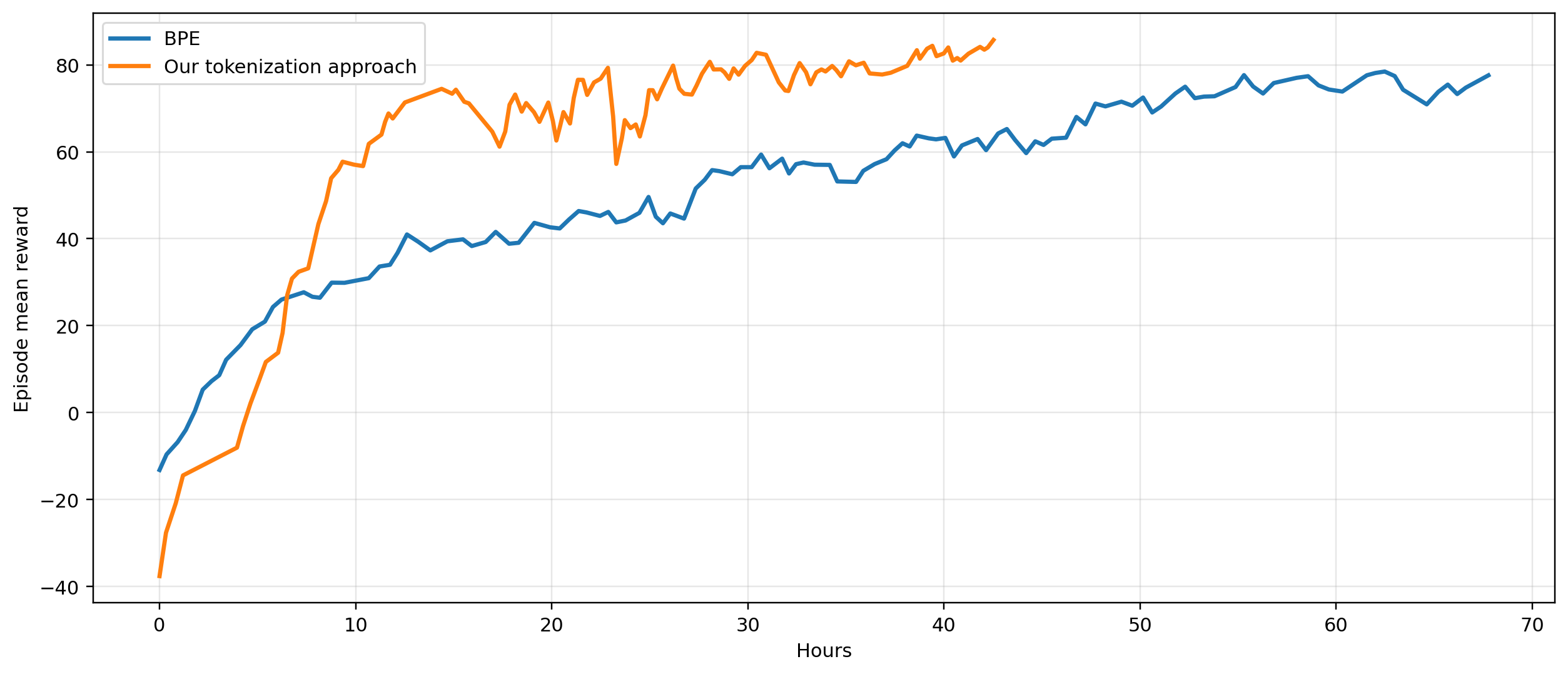}
    \shrink[2]
    \caption{Graph comparing Episode Mean Reward Over Training Time of the CHEHAB RL with \Token{} tokenization and CHEHAB RL with BPE (both trained for 2 million steps).}
    \label{fig:training_time_bpe_vs_dynamic}
    \Description{A plot showing the training curves of two models over 2 million steps. The y-axis represents the Episode Mean Reward and the x-axis represents training time. It compares the performance of CHEHAB RL using the custom tokenization versus Byte Pair Encoding (BPE).}
    \shrink[1]
\end{figure}
\input{tables/kernel}
\ifshowextra

\subsection{GRU vs. Transformer Encoder}
\label{appendix:ablation_encoder}

To validate our choice of a Transformer encoder for the state representation, we conducted an experiment to compare its encoding capabilities against a baseline recurrent architecture, the Gated Recurrent Unit (GRU). For this comparison, we constructed complete autoencoders for both architectures. The quality of the reconstruction serves as a direct measure of the quality of embedding learned by the encoder; a perfect reconstruction implies that the encoder has preserved all necessary structural information of the program IR. Our RL agent uses only the encoder component, but this autoencoder setup allows us to empirically verify its ability to learn an embedding for the input program.

\paragraph{Experimental Setup}
Both autoencoders were trained on the same dataset of 1.4 million randomly generated IR expressions. We use the same methodology described in Sec.~\ref{randomdata} to generate this dataset. The GRU baseline used a 4-layer bidirectional encoder and a 4-layer decoder. The Transformer autoencoder also used 4 encoder and 4 decoder layers, with both models using identical optimizer settings and batch sizes.

\paragraph{Results}
The training curves in Fig.~\ref{fig:ae_learning_curves_appendix} and the final test results in Table~\ref{tab:ae_ablation_appendix} show a clear distinction in performance. The Transformer autoencoder not only learns significantly faster but also achieves 100\% exact-match reconstruction accuracy on the test set. The GRU model, however, plateaus at 98.92\% exact-match accuracy.

\begin{figure}[h!t]
  \centering
  \includegraphics[width=0.8\linewidth]{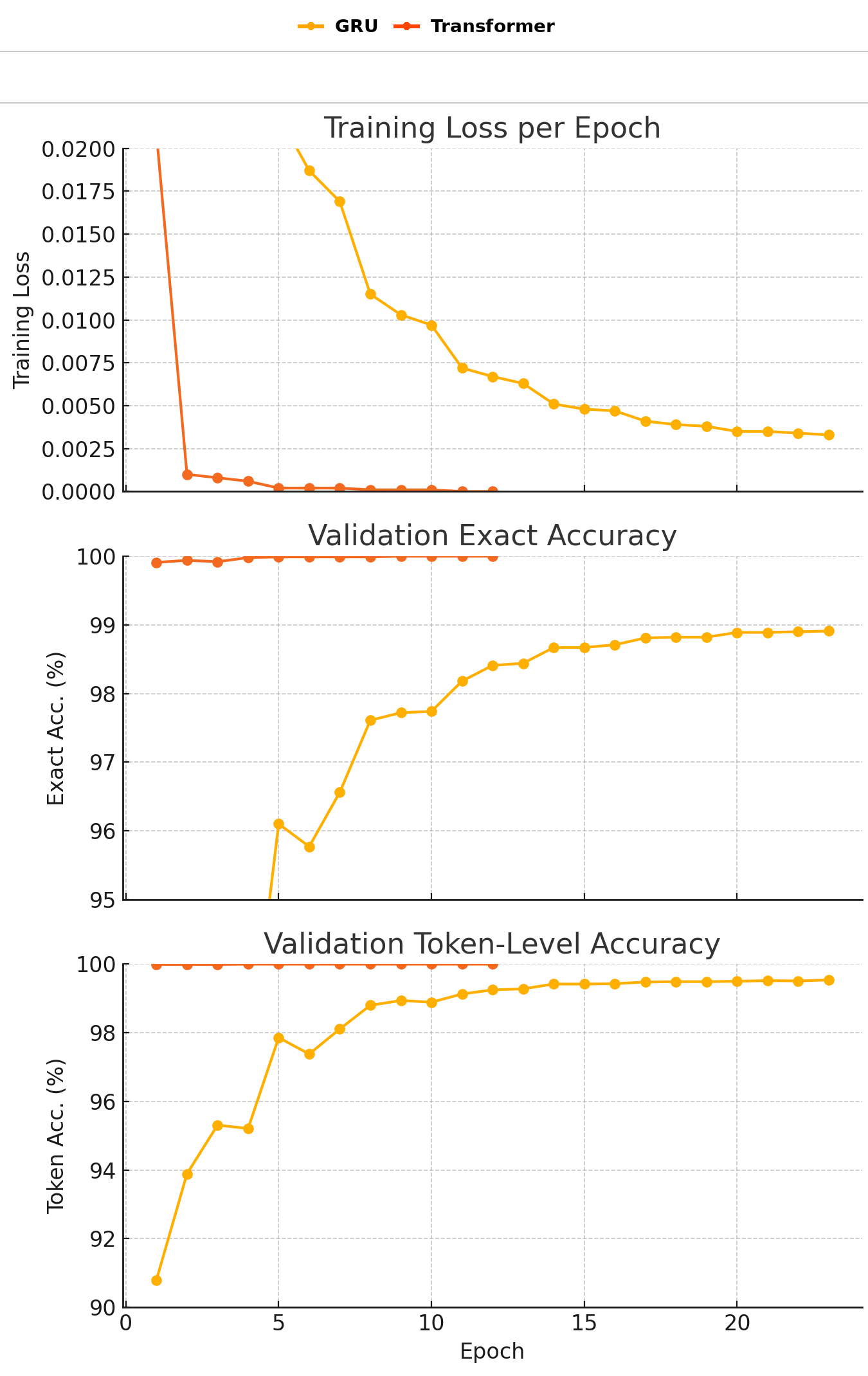}
  \caption{Validation loss and accuracy comparison between the Transformer and GRU-based autoencoders. RNN-AE in the figure represents the GRU-based auto-encoder.}
  \Description{Two plots comparing the training progress of Transformer and GRU-based autoencoders.}
  \label{fig:ae_learning_curves_appendix}
\end{figure}

\begin{table}[H]
  \CodeSize
  \centering
  \caption{Autoencoder reconstruction accuracy on the validation and test set.}
  \label{tab:ae_ablation_appendix}
  \begin{tabular}{@{}lccccc@{}}
    \toprule
    \multirow{2}{*}{\textbf{Model}} &
      \multicolumn{2}{c}{\textbf{Validation}} &
      \phantom{ab} &
      \multicolumn{2}{c}{\textbf{Test}} \\
    \cmidrule(r){2-3}\cmidrule(l){5-6}
                  & Exact (\%) & Token (\%) && Exact (\%) & Token (\%) \\
    \midrule
    GRU Autoencoder      & 98.91 & 99.54 && 98.92 & 99.52 \\
    Transformer Autoenc. & \textbf{100.00} & \textbf{100.00} &&
                           \textbf{100.00} & \textbf{100.00} \\
    \bottomrule
  \end{tabular}
\end{table}

\paragraph{Analysis of Results}
The primary source of failure for the GRU was ordering errors, where it produced the correct set of tokens but in an incorrect order. A detailed analysis showed that 9.8\% of the GRU's reconstructions had a non-zero edit distance from the ground truth, typically due to misplaced parentheses or swapped sibling nodes in the expression tree. This confirms the hypothesis that the GRU's sequential processing and fixed-length hidden state are insufficient to reliably preserve the long-range structural information of complex IRs. The Transformer's self-attention mechanism, in contrast, processes all tokens in parallel, allowing it to directly model these dependencies and achieve lossless reconstruction. These results validate our choice of the Transformer encoder for state representation.

\fi

\ifshowextra 

\section{More Detailed Related Work\label{appendix:related}}

\paragraph{FHE Compilers for Circuit Optimization}

EVA~\cite{Dathathri2020} is an example of such a compiler, it focuses on automatic parameters selection and ciphertext maintenance to manage noise and scale of ciphertexts for the CKKS scheme, however, it does not apply any optimization strategies in order to reduce the multiplicative depth or the number of operations. In a later version of the same work \cite{Chowdhary2021EVA}, CSE (Common Subexpression Elimination) was added, which provides basic simplifications of the circuit (in CSE, redundant expressions are computed once and used multiple times). CHEHAB also applies CSE on the generated code, but in addition to that, it applies term rewriting to reduce the number of operations and the multiplicative depth. Ramparts~\cite{Archer2019} uses Julia as input language and targets the BFV scheme. It provides automatic parameter selection based on noise growth estimation, it also simplifies circuits in order to reduce the number of operations by applying simple and classical optimization techniques such as CSE and constant folding, it also offers optimizations such as loop unrolling, and function inlining. It does insert ciphertext-maintenance operation, but in a naive way \cite{Viand2021Sok}. Unlike CHEHAB, Ramparts cannot optimize vectorized code~\cite{Viand2021Sok}, which affects its ability to take advantage of SIMD parallelism and improve latency.
HECATE~\cite{Lee2022} is a recent work that focuses on selecting the parameter \(q\) for CKKS, and at the same time it performs ciphertext-maintenance operations scheduling by applying rewrite on the input circuit, HECATE outperforms EVA by \(27.38\%\), however, it does not apply any optimization to reduce the number of operations and multiplicative depth of the circuit.

\paragraph{RL-based methods for Code Optimization.}

Recent attempts explored the use of reinforcement learning to solve the problem of choosing the right sequence of code transformations. In PolyGym~\cite{brauckmann2021reinforcement} and CompilerGym~\cite{cummins2022compilergym}, the authors propose only RL environments without implementing RL agents to optimize code; their main contribution is to show that their action space has potentially good optimizations to explore. They leave the implementation of an RL agent as future work.

Other work such as HalideRL~\cite{pecenin2019optimization}, AutoPhase~\cite{huang2020autophase} and SuperSonic~\cite{huanting2022ss} propose RL agents to optimize code.
HalideRL is not fully automatic. The user has to provide an initial set of code transformations. The HalideRL agent then discards transformations that are not useful and keeps only those that are useful. It then selects the best parameters for the useful transformations. In addition, HalideRL does not generalize to programs unseen during training. It is trained on a given program with multiple random data input sizes. Then, during deployment, it is used to optimize that same program. This is different from our approach. Our RL agent is designed to generalize to programs unseen during training. We train our RL agent on a large set of random LLM-generated programs. Once it learns how to optimize them, we then deploy it on new unseen programs and use it to optimize them.

SuperSonic~\cite{huanting2022ss} is a meta-optimizer that targets the problem of choosing the best RL algorithm and the best representation of states and actions, while AutoPhase~\cite{huang2020autophase} targets the problem of phase ordering, i.e., selecting the best order for compiler passes.

While previous work addresses traditional code optimizations, it does not consider the unique constraints of FHE. Our work is the first to formulate FHE optimization as an RL problem, where the agent must learn a policy that navigates the trade-offs between vectorization, cryptographic noise accumulation, and the cost of operations.

\paragraph{Tokenization and Variable Renaming}

Tokenization has been widely used in NLP. Classical NLP moved from word-level tokenization to learned subword units, with methods such as \emph{Byte-Pair Encoding} (BPE) \cite{sennrich-etal-2016-neural}; more recently, tokenization-free methods process raw characters or bytes (e.g., \emph{CANINE} \cite{Clark_2022}), which eliminates vocabulary design but lengthens sequences and shifts compute into the model. In coding tasks, BPE has been widely used as a tokenization method for models that take code as input in its textual format (e.g. \emph{CodeBERT~\cite{feng2020codebert}}). None of these methods applies variable renaming at the tokenization level, though.

Variable renaming to obtain canonical representations of code was explored for classical clone detection, as in \emph{CCFinder} \cite{kamiya2002ccfinder}. Such variable renaming was not used as a tokenization method for deep learning models, though.

Our approach differs from the previous approaches. We normalize the IR by replacing variable names and numeric constants with generic tokens, since their exact values do not affect our rewrite rules, which means that our tokenization method is more specific to our use case. This makes it faster to learn than BPE but still preserves the structural signals the model needs.

\paragraph{Random Code Generation}

Random code generation is a common strategy for constructing training corpora for learning-based compiler optimization. Systems such as the Tiramisu autoscheduler~\cite{baghdadi2021deeplearningbasedcost}, Looper~\cite{merouani2025looperlearnedautomaticcode}, and the Halide autoscheduler~\cite{LearningHalide} use stochastic code generators in which the probabilities of syntactic and semantic patterns are manually specified to sample diverse code patterns. To eliminate this manual specification burden, Cummins et al.~\cite{7863731} learn a generative model from large-scale real-world code and then sample synthetic programs whose distribution better matches human-written code. More recently, PIE~\cite{shypula2024learningperformanceimprovingcodeedits} and EffiCoder~\cite{huang2025efficoderenhancingcodegeneration} use large language models (LLMs) to synthesize code that is subsequently used to train or fine-tune downstream models for performance-oriented code optimization. Our approach follows this latter line: we leverage an off-the-shelf, pretrained LLM, already exposed to the distribution of real-world code, to synthesize training programs, thereby avoiding task-specific generator design and hand-tuning of pattern priors.

\fi

\begin{figure*}[t]
    \vspace{0.3cm}
    \centering
    \includegraphics[width=\textwidth]{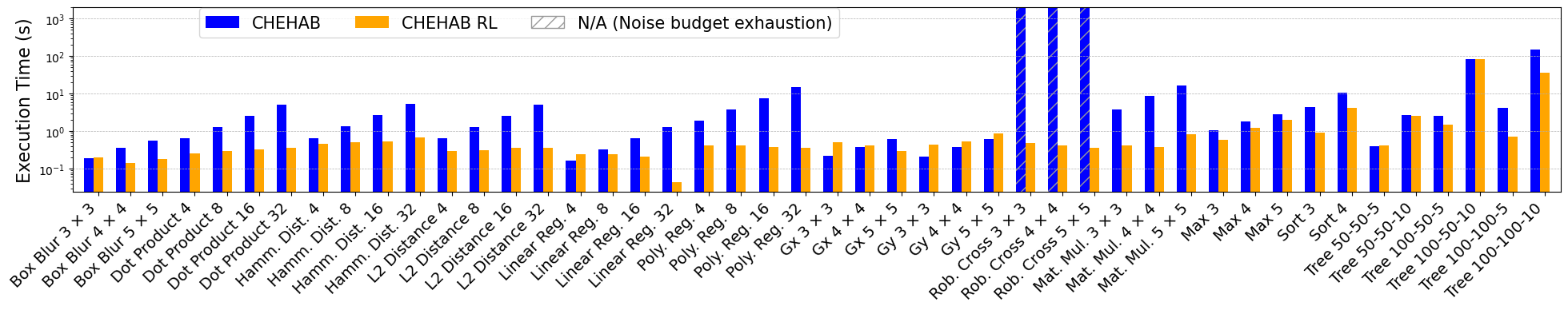}
    \shrink[3]
    \caption{Semi-log plots comparing the execution times of the default CHEHAB and CHEHAB RL.}
    \label{fig:execution_time_chehab_vs_rl}
    \Description{Semi-logarithmic plots displaying the execution time performance of the default CHEHAB compiler compared to the CHEHAB RL approach.}
    \shrink[1]
\end{figure*}

\ifshowextra
\begin{figure}[t]
    \centering
    \includegraphics[width=\linewidth]{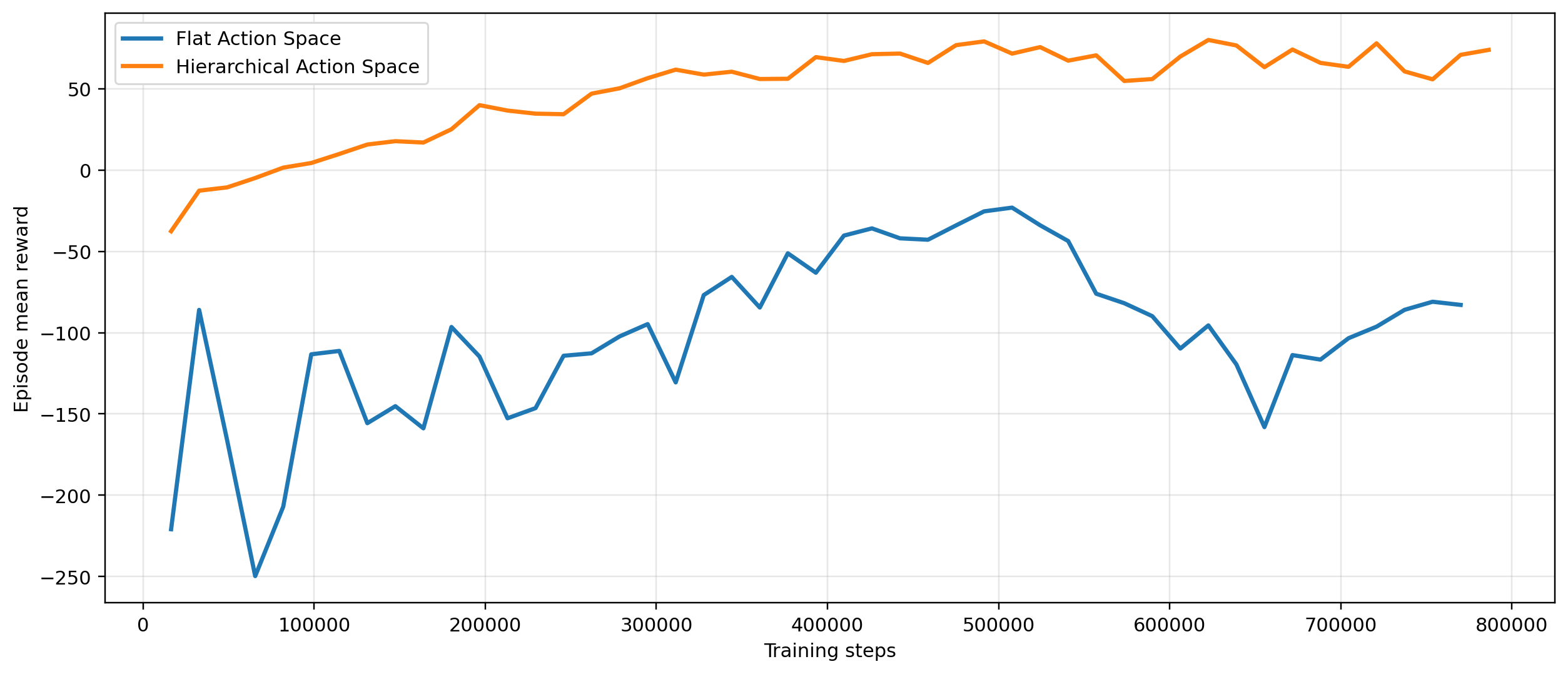}
    \shrink[2]
    \caption{Graph comparing Episode Mean Reward Over Timesteps of Flat vs. Hierarchical Action Spaces.}
    \Description{A comparison graph of reinforcement learning performance. The y-axis shows Episode Mean Reward and the x-axis shows Timesteps. The plot contrasts the learning efficiency of a Flat Action Space against a Hierarchical Action Space.}
    \label{fig:flat_vs_hier}
    \shrink[1]
\end{figure}
\fi

\ifshowextra 
\section{Frequently Asked Questions}

In this section, we provide a list of frequently asked questions and answers to these questions. These provide more clarification about the paper.

\begin{itemize}
    \item \textbf{Examples of end-user deployments for homomorphic encryption}
    End‑user deployments for homomorphic encryption have already appeared: for example, Microsoft Edge’s Password Monitor uses homomorphic encryption to check user credentials against breach corpora without revealing inputs~\cite{MSR2021PasswordMonitor,Viand2022HECOFH}. Beyond credentials checking, homomorphic encryption is being used in sensitive scientific/health settings (e.g., private genotype imputation and kinship detection in genomics), both reporting practical runtimes on real datasets~\cite{gursoy2022privacy,KIM20211108}. The UK Information Commissioner’s Office documents a cross‑institution deployment in which banks and law‑enforcement agencies use homomorphic encryption to run encrypted queries to detect financial crime~\cite{ICO2023HECaseStudy}.
    
    \item  \edit{\textbf{Why not use an LLM to directly optimize CHEHAB IR at inference time?}
    It is an interesting direction. In this work, our goal is \emph{fast and deterministic} compilation, which favors an RL policy evaluated locally inside the compiler. At inference time, the RL agent produces a rewrite sequence in a few seconds with stable runtime and fully reproducible behavior. In contrast, LLM-based optimization at inference time typically incurs substantially higher latency and cost, and can be less predictable due to sampling and the need for output validation/repair. For example, the ComPilot paper \cite{2025compilot} reports an average of 8 minutes to optimize a single loop nest.}

\end{itemize}
\fi

\section{Artifact Appendix}
\subsection{Abstract}
CHEHAB is a fully homomorphic encryption (FHE) compiler that translates a domain-specific language into Microsoft SEAL (BFV) programs and applies optimization passes including a reinforcement-learning-guided rewrite selection, constant folding and common subexpression elimination. The artifact reproduces the key results of our paper (Table \ref{tab:kernel}, Figure \ref{fig:kernel}) by running the benchmark suite (e.g., Box Blur, Dot Product, Hamming Distance, L2 Distance, Linear/Polynomial Regression, Matrix Multiplication, Max, Sort, and polynomial-tree benchmarks) under the same optimization configurations used in the evaluation, and collecting compile-time and execution-time metrics as well as circuit-level properties (depth, multiplicative depth, remaining noise budget, and operation counts). The workflow produces CSV result files in \path{results/} and includes scripts to generate the corresponding plots from these CSVs. The artifact is packaged with a ready-to-use Docker environment (Ubuntu + SEAL + Conda dependencies) for interactive use of the compiler, and an optional web interface for running individual benchmark configurations and inspecting logs and optimized expressions.

\shrink[1]
\subsection{Artifact check-list (meta-information)}
{\small
\begin{itemize}
  \item {\bf Algorithm:} RL-guided optimization for FHE compilation (learned rewrite selection)
  \item {\bf Program:} C++ compiler + benchmarks, Python benchmark driver, optional FastAPI web UI
  \item {\bf Compilation:} CMake, GCC/G++, Microsoft SEAL (BFV)
  \item {\bf Transformations:} RL-guided rewrite selection, constant folding, common subexpression elimination
  \item {\bf Binary:} benchmark executables in \path{build/benchmarks/<bench>/} HE runner \texttt{main} in generated \path{he/}
  \item {\bf Model:} RL agent code under \path{RL/fhe\_rl/} (trained models included)
  \item {\bf Data set:} LLM-generated dataset used for training (no external dataset)
  \item {\bf Run-time environment:} Docker (Ubuntu 22.04) + Conda env \path{chehabEnv}; optional host install per README
  \item {\bf Hardware:} x86\_64 CPU, $\ge$32\,GB RAM recommended, for our experiments we used 1 node, 32 cores, 128\,GB RAM
  \item {\bf Execution:} \path{python run\_benchmarks.py} (RL mode); optional \path{docker compose up chehab-demo} for web UI
  \item {\bf Metrics:} compile time, execution time, depth, multiplicative depth, remaining noise budget, operation counts
  \item {\bf Output:} CSV files in \path{results/}; optional plots (PNG) from \texttt{results/generate\_graphs.py}
  \item {\bf Experiments:} benchmark suite runs across multiple slot counts and RL settings; per-benchmark runs via CLI or web UI
  \item {\bf How much disk space required (approximately)?:} 25-30\,GB (Docker image + build + outputs)
  \item {\bf How much time is needed to prepare workflow (approximately)?:} 45-60 minutes (first Docker build)
  \item {\bf How much time is needed to complete experiments (approximately)?:} 45-90 minutes (depends on iterations)
  \item {\bf Publicly available?:} Yes
  \item {\bf Workflow automation framework used?:} Python scripts, Docker Compose
\end{itemize}
}
\subsection{Description}

\subsubsection{How to access}
The artifact is available as a public repository: \url{https://github.com/Modern-Compilers-Lab/CHEHAB}. 
A Docker-based environment is provided for reproducible setup and execution. A host-native installation is also possible by following the repository \texttt{README} (including Microsoft SEAL and toolchain installation).

\subsubsection{Software dependencies}
Recommended: Docker and Docker Compose.
The Docker image includes Ubuntu 22.04, Microsoft SEAL, the required build toolchain, and a Conda environment (\texttt{chehabEnv}) with Python dependencies used by the RL optimization pipeline.

\subsection{Installation}

\paragraph{Build the interactive environment.}
From the repository root, build the interactive environment:
\begin{lstlisting}[style=shell]
$ docker compose build chehab-main
\end{lstlisting}

\paragraph{Build the web UI service (optional).}
From the repository root, build the web UI service:
\begin{lstlisting}[style=shell]
$ docker compose build chehab-demo
\end{lstlisting}

\paragraph{Host setup (optional).}
If you prefer running without Docker, follow the \texttt{README} for the complete host installation procedure (including Microsoft SEAL and all required dependencies).

\subsection{Experiment workflow}

\paragraph{CLI workflow (recommended).}
Launch an interactive container shell:
\begin{lstlisting}[style=shell]
$ docker compose run --rm -it \
  chehab-main /bin/bash
\end{lstlisting}
The Docker image is configured to auto-activate \texttt{chehabEnv} for interactive \texttt{bash} shells. If it is not active, run:
\begin{lstlisting}[style=shell]
$ source /opt/conda/etc/profile.d/conda.sh
$ conda activate chehabEnv
\end{lstlisting}
Run the benchmark suite:
\begin{lstlisting}[style=shell]
$ python run_benchmarks.py
\end{lstlisting}
The script writes the results as CSV files in \path{results/}, this directory is bind-mounted, so the files are mirrored to the local \path{results/} folder on the host.

\paragraph{Web workflow (optional).}
Start the web service:
\begin{lstlisting}[style=shell]
$ docker compose up chehab-demo
\end{lstlisting}
Then open \url{http://localhost:8000} to run individual benchmark configurations and inspect logs.

\subsection{Evaluation and expected results}
Successful execution produces CSV result files in \path{results/}. Each row corresponds to a benchmark configuration and reports compile-time and execution-time measurements, along with circuit-level statistics (e.g., depth, multiplicative depth, remaining noise budget, and operation counts). Optional plots can be generated from the CSV outputs using \path{generate_graphs.py}.
a file such as \path{results/results_RL.csv} is produced. To generate an execution-time plot:

\begin{lstlisting}[style=shell]
$ python results/generate_graphs.py \
  --metric exec \
  --csv results/results_RL.csv \ 
  --label "CHEHAB RL" \
  --output results/exec_time.png
\end{lstlisting}

To plot remaining noise budget:

\begin{lstlisting}[style=shell]
$ python results/generate_graphs.py \
  --metric noise \
  --csv results/results_RL.csv \
  --label "CHEHAB RL" \
  --output results/noise_budget.png
\end{lstlisting}

\subsection{Experiment customization}
The benchmark sweep parameters (e.g., slot counts, number of iterations, and timeouts) can be modified in \path{run_benchmarks.py}. When using Docker Compose, \path{run_benchmarks.py} is bind-mounted into the container, so changes are mirrored immediately and do not require rebuilding the image. The web UI exposes common parameters directly in its input form.

RL-specific customization is done in \path{RL/fhe_rl/config.py}, which selects the RL model to use, for example, users can switch to the agent trained with the random dataset by changing the configured model path. This file is also bind-mounted in the Docker workflow, so edits take effect without rebuilding.

\subsection{Notes}
The Docker workflow is recommended to avoid dependency and toolchain mismatches. If running outside Docker, follow the README to install Microsoft SEAL and all dependencies.

%% file: tables/notation.tex
\begin{table}
    \fontsizefortables
    \centering
    \caption{Notation.}
    \shrink
    \label{tab:notation}
    \begin{tabular}{c|l}
        \hline
        Symbol            & Description \\ \hline
        $\cup$            & depth including all operations \\
        $\cup^\otimes$    & multiplicative depth \\ 
        $\oplus$          & \# of ciphertext additions \\
        $\otimes$         & \# of ciphertext-ciphertext multiplications \\
        $\odot$           & \# of ciphertext-plaintext multiplications \\
        $\boxtimes$       & \# of ciphertext squares \\
        $\circlearrowright$ & \# of rotations \\ \hline
    \end{tabular}
    \shrink[1]
\end{table}

%% file: tables/kernel.tex
\setlength\tabcolsep{2.2pt}
\setlength\arrayrulewidth{0.8pt}
\begin{table*}[h!t]
    \vspace{0.2cm}
    \fontsizefortables
    \centering
    \caption{
    Comparison between four configurations: 1) an initial, naive, implementation of the benchmarks; 2) CHEHAB RL; 3) Coyote; 4) CHEHAB RL with data layout transformation applied after encryption. We compare them in terms of circuit depth ($\cup$) and multiplicative depth ($\cup^\otimes$), number of ciphertext-ciphertext multiplications ($\otimes$), number of rotations ($\circlearrowright$), number of ciphertext-plaintext multiplications ($\odot$), and number of ciphertext additions ($\oplus$). We also report CHEHAB's and Coyote's compilation times CT (s) and their Consumed Noise (CN). The important columns in the Table are in bold.}
    \label{tab:kernel}


\begin{tabular}{l|cccccc|cccccccc|cccccccc|ccccccc}
        \hline
        \multirow{2}*{Kernel}
        & \multicolumn{6}{c|}{Initial}
        & \multicolumn{8}{c|}{CHEHAB RL}
        & \multicolumn{8}{c|}{Coyote}
& \multicolumn{7}{c}{\makecell{CHEHAB RL\\ with data layout transformed \\ after encryption}} \\        {}
        &  $\cup$ &     $\cup^\otimes$  & $\otimes$ &     $\circlearrowright$  & $\odot$ & $\oplus$
        &  $\cup$ & \h{$\cup^\otimes$}  & $\otimes$ & \h{$\circlearrowright$} & $\odot$ & $\oplus$ & \h{CN} & \h{CT}
        &  $\cup$ & \h{$\cup^\otimes$}  & $\otimes$ & \h{$\circlearrowright$} & $\odot$ & $\oplus$ & \h{CN} & \h{CT}
        &  $\cup$ & \h{$\cup^\otimes$}  & $\otimes$ & \h{$\circlearrowright$} & $\odot$ & $\oplus$ & \h{CT}
        \\ \hline
Box Blur 3 × 3&9&0&0&0&0&31&6&0&0&1&0&6&13.0&9.57&25&0&0&18&45&24&141.0&471.991&16&1&30&13&30&35&10.568\\
Box Blur 4 × 4&9&0&0&0&0&74&5&0&0&1&0&5&12.0&10.146&36&0&0&51&111&26&247.0&926.780&15&1&31&13&31&34&9.997\\
Box Blur 5 × 5&9&0&0&0&0&135&4&0&0&0&0&8&10.0&10.457&36&0&0&100&204&27&250.0&1553.897&10&1&29&8&29&33&10.129\\
\hline
Dot Product 4&5&1&4&0&0&4&6&1&1&3&0&2&45.0&9.826&7&1&1&2&4&4&42.0&106.571&13&2&7&8&6&7&9.21\\
Dot Product 8&9&1&8&0&0&8&8&1&1&4&0&3&47.0&9.864&9&1&1&3&4&6&93.0&179.196&15&2&7&9&6&8&9.417\\
Dot Product 16&17&1&16&0&0&16&10&1&1&5&0&4&47.0&11.068&15&1&1&10&14&11&119.0&324.983&29&2&19&22&18&21&11.162\\
Dot Product 32&33&1&32&0&0&32&11&1&1&5&0&5&44.0&20.492&22&1&1&21&34&16&145.0&629.485&21&2&19&21&18&21&20.536\\
\hline
Hamm. Dist. 4&7&2&8&0&4&8&8&2&2&4&0&3&74.0&9.421&3&2&2&0&0&1&99.0&144.852&14&3&16&12&14&16&9.319\\
Hamm. Dist. 8&11&2&16&0&8&16&10&2&2&5&0&4&74.0&11.416&4&2&2&3&0&1&73.0&282.906&16&3&16&13&14&17&11.365\\
Hamm. Dist. 16&19&2&32&0&16&32&12&2&2&6&0&5&75.0&29.736&12&2&2&21&42&12&233.0&520.420&30&3&40&30&38&42&29.816\\
Hamm. Dist. 32&35&2&64&0&32&64&14&2&3&5&0&7&76.0&178.241&12&2&2&69&92&8&156.0&1054.597&24&3&41&21&38&43&176.251\\
\hline
L2 Distance 4&5&1&4&0&0&3&8&1&1&4&0&2&45.0&9.445&8&1&1&2&4&4&68.0&158.341&21&2&13&15&12&13&9.321\\
L2 Distance 8&9&1&8&0&0&7&9&1&1&4&0&3&45.0&10.35&11&1&2&3&9&9&122.0&276.027&16&2&13&14&12&13&10.351\\
L2 Distance 16&17&1&16&0&0&15&11&1&1&5&0&4&46.0&20.696&22&1&2&23&52&18&176.0&574.150&28&2&33&35&32&34&20.735\\
L2 Distance 32&33&1&32&0&0&31&12&1&1&5&0&5&45.0&95.935&26&1&1&73&101&17&179.0&1170.184&22&2&19&21&18&21&95.549\\
\hline
Linear Reg. 4&3&1&4&0&4&8&5&1&1&2&0&2&41.0&9.36&3&1&1&0&0&0&41.0&2.278&8&2&3&3&2&4&8.995\\
Linear Reg. 8&3&1&8&0&8&16&5&1&1&2&0&2&41.0&9.469&6&1&1&1&6&3&66.0&197.718&8&2&3&3&2&4&9.312\\
Linear Reg. 16&3&1&16&0&16&32&4&1&1&1&0&2&41.0&9.547&12&1&1&17&25&7&122.0&500.449&7&2&3&2&2&4&9.283\\
Linear Reg. 32&3&1&32&0&32&64&3&1&1&0&1&2&10.0&9.778&12&1&1&44&46&4&124.0&963.687&4&2&3&0&3&2&9.466\\
\hline
Poly. Reg. 4&5&2&12&0&0&12&8&2&2&3&0&3&74.0&10.139&16&2&3&4&19&10&153.0&238.869&14&3&8&8&6&9&9.955\\
Poly. Reg. 8&5&2&24&0&0&24&7&2&2&2&0&3&73.0&9.786&20&2&2&21&38&12&205.0&501.163&10&3&5&4&3&5&9.694\\
Poly. Reg. 16&5&2&48&0&0&48&6&2&2&1&0&3&73.0&9.993&20&2&2&56&79&12&208.0&961.519&9&3&7&5&5&7&10.06\\
Poly. Reg. 32&5&2&96&0&0&96&5&2&2&0&0&3&73.0&10.145&20&2&2&134&173&12&210.0&1896.952&6&3&7&0&5&3&9.952\\
\hline
Gx 3 × 3&4&1&24&0&24&36&8&2&4&1&2&6&66.0&9.463&16&1&1&27&57&11&150.0&594.926&12&3&18&9&16&20&9.363\\
Gx 4 × 4&4&1&48&0&48&64&6&1&3&1&1&4&42.0&10.743&16&1&1&59&112&11&150.0&1061.439&12&2&16&9&14&17&10.599\\
Gx 5 × 5&4&1&80&0&80&100&5&1&3&0&2&5&40.0&10.646&16&1&1&105&182&12&154.0&1673.958&9&2&16&6&15&16&10.675\\
\hline
Gy 3 × 3&4&1&24&0&24&33&6&1&3&1&1&3&41.0&9.68&16&1&1&29&54&10&149.0&595.976&14&2&20&11&18&20&9.737\\
Gy 4 × 4&4&1&48&0&48&64&6&1&3&1&0&3&42.0&10.591&16&1&1&60&101&11&197.0&1057.486&14&2&23&11&20&23&10.988\\
Gy 5 × 5&4&1&80&0&80&105&6&1&5&0&0&6&42.0&11.322&17&1&1&103&169&11&154.0&1670.887&11&2&25&6&20&26&11.197\\
\hline
Rob. Cross 3 × 3&6&2&42.0&0.0&24.0&63.0&5&1&2&2&0&3&45.0&9.786&12&1&2&10&27&11&96.0&291.833&11&2&12&9&10&12&9.661\\
Rob. Cross 4 × 4&6&2&76.0&0.0&44.0&112.0&5&1&2&2&0&3&44.0&10.363&12&1&2&24&51&9&125.0&543.704&11&2&12&9&10&12&10.133\\
Rob. Cross 5 × 5&6&2&120.0&0.0&70.0&175.0&4&1&2&0&0&4&42.0&11.681&12&1&2&43&95&10&126.0&857.648&8&2&9&3&7&10&11.505\\
\hline
Mat. Mul. 3 × 3&3&1&27&0&0&18&4&1&2&1&0&2&41.0&9.426&9&1&2&9&14&5&93.0&456.644&11&2&19&11&17&17&9.452\\
Mat. Mul. 4 × 4&4&1&64&0&0&48&4&1&2&1&0&2&42.0&10.039&10&1&2&32&44&7&94.0&1085.128&13&2&30&16&28&26&9.952\\
Mat. Mul. 5 × 5&5&1&125&0&0&100&4&1&5&0&0&4&42.0&11.601&14&1&1&69&82&13&145.0&2158.145&10&2&55&16&50&44&11.324\\
\hline
Max 3&6&2&6&0&0&6&9&2&3&3&0&4&76.0&9.481&14&2&3&3&10&9&75.0&132.890&9&2&3&3&0&4&9.481\\
Max 4&8&3&12&0&0&12&11&3&7&3&0&8&107.0&11.784&22&3&6&8&25&16&235.0&252.436&11&3&7&3&0&8&11.784\\
Max 5&10&4&20&0&0&20&15&4&12&5&0&12&140.0&149.936&29&4&9&27&55&21&321.0&509.189&15&4&12&5&0&12&149.936\\
\hline
Sort 3&8&3&10&0&0&8&11&3&5&3&0&6&107.0&10.552&12&3&4&5&6&9&188.0&184.944&11&3&5&3&0&6&10.552\\
Sort 4&14&6&58&0&0&35&21&6&26&5&0&35&206.0&522.043&42&6&10&66&114&31&369.0&959.704&21&6&26&5&0&35&522.043\\
\hline
Tree 50-50-5&4&2&2.0&0.0&0.0&4.0&8&2&2&3&0&3&73.0&9.271&4&2&2&0&0&4&73.0&4.306&8&2&2&3&0&3&9.271\\
Tree 50-50-10&10&5&16.0&0.0&0.0&16.0&15&5&15&2&0&18&172.0&18.17&27&7&7&7&19&21&243.0&147.033&15&5&15&2&0&18&18.17\\
Tree 100-50-5&5&3&15.0&0.0&0.0&16.0&13&5&8&3&0&13&169.0&12.428&18&3&4&7&22&14&212.0&176.904&13&5&8&3&0&13&12.428\\
Tree 100-50-10&10&8&519.0&0.0&0.0&504.0&16&8&509&2&0&511&334.0&148.398&50&8&148&490&873&243&369.0&6798.692&16&8&509&2&0&511&148.398\\
Tree 100-100-5&5&4&29.0&0.0&0.0&2.0&10&4&4&5&0&1&141.0&11.152&16&5&5&5&13&7&211.0&170.154&10&4&4&5&0&1&11.152\\
Tree 100-100-10&10&9&1022.0&0.0&0.0&1.0&12&9&256&2&0&1&307.0&91.627&36&10&10&334&373&16&369.0&5396.742&12&9&256&2&0&1&91.627\\

   \hline
    \end{tabular}
    
\end{table*}